\documentclass{article}

\usepackage[preprint]{neurips_2025}

\usepackage[table]{xcolor}
\usepackage{threeparttable}
\usepackage{color}
\usepackage{colortbl}
\usepackage{nicematrix}
\usepackage{array}
\usepackage{booktabs}
\usepackage{graphicx}
\usepackage{makecell}
\usepackage{pifont}            
\usepackage{multirow}
\newcommand{\cmark}{\ding{51}} 
\newcommand{\xmark}{\ding{55}} 
\usepackage{tikz}
\usepackage{amsmath}
\usepackage{markdown}

\usepackage[linesnumbered,ruled,vlined]{algorithm2e}
\usepackage[most]{tcolorbox}
\usepackage{float}
\usepackage{fontawesome5}

\definecolor{promptcolor}{HTML}{6E8EC9}
\tcbset{
  fancyprompt/.style={
    enhanced,
    breakable, 
    colback=promptcolor!5!white,
    colbacklower=promptcolor!10!white,
    colframe=promptcolor,
    coltitle=white,
    fonttitle=\bfseries\footnotesize,
    boxrule=0.5pt,
    arc=6pt,
    drop shadow southeast,
    attach boxed title to top left={yshift=-2mm, xshift=2mm},
    boxed title style={
      colback=promptcolor!75!black,
      colframe=promptcolor!75!black,
      rounded corners
    },
    top=10pt,
    bottom=10pt,
    left=10pt,
    right=10pt
  }
}

\usepackage{listings}

\definecolor{codegreen}{rgb}{0,0.6,0}
\definecolor{codegray}{rgb}{0.5,0.5,0.5}
\definecolor{codepurple}{rgb}{0.58,0,0.82}
\definecolor{backcolour}{rgb}{0.95,0.95,0.92}

\lstdefinestyle{mystyle}{
    commentstyle=\color{codegreen},
    keywordstyle=\color{magenta},
    numberstyle=\tiny\color{codegray},
    stringstyle=\color{codepurple},
    basicstyle=\footnotesize\ttfamily,
    breaklines=true,
    captionpos=b,
    keepspaces=true,
    numbers=left,
    numbersep=5pt,
    showspaces=false,
    showstringspaces=false,
    showtabs=false,
    tabsize=2
}

\definecolor{AlgBlue}{RGB}{0,85,164}
\definecolor{AlgRed}{RGB}{239,65,53}
\definecolor{AlgGreen}{RGB}{100,161,56}
\definecolor{ForestGreen}{rgb}{0.13, 0.55, 0.13}

\SetKwComment{Comment}{\color{AlgGreen}\% }{}

\SetKwProg{Function}{function}{}{}

\usepackage{booktabs}  
\usepackage{makecell}  
\usepackage{acronym}
\usepackage{mathtools}

\long\def\comment#1{}
\usepackage{wrapfig}
\usepackage{textcomp,booktabs}
\usepackage{multirow}
\usepackage{rotating}
\usepackage{epstopdf}
\usepackage{xspace}
\usepackage{bbm}
\usepackage[short]{optidef}
\definecolor{Gray}{HTML}{EAECF1}
\definecolor{Brown}{HTML}{964B00}
\definecolor{Green}{HTML}{70BE71}
\definecolor{BlueGray}{HTML}{94A3AD}
\definecolor{White}{RGB}{255,255,255}
\usepackage{comment}
\usepackage{bbold}
\usepackage[labelformat=simple, subrefformat=parens, textfont=normal, labelfont=normal, list=true]{subcaption}

\usepackage{enumitem}
\usepackage{lipsum}
\usepackage{soul}
\usepackage[normalem]{ulem}
\usepackage[usestackEOL]{stackengine}
\usepackage{listings}
\usepackage[T1]{fontenc}
\usepackage{fancyvrb}

\newcommand{\otoprule}{\midrule[\heavyrulewidth]}

\newcommand{\eg}{{\it e.g.}}
\newcommand{\etal}{\textit{et al}.~}

\newcommand{\sys}{\textit{\textbf{CodeMirage}}\xspace}

\newcommand{\paraspace}{\vspace{0.05in}}
\newcommand{\parab}[1]{\paraspace\noindent{\bf #1}}

\usepackage{amsthm}

\usepackage{enumitem}
\setlist[description]{leftmargin=\parindent,labelindent=\parindent}

\usepackage[utf8]{inputenc} 
\usepackage[T1]{fontenc}    
\usepackage{hyperref}       
\usepackage{url}            
\usepackage{booktabs}       
\usepackage{amsfonts}       
\usepackage{nicefrac}       
\usepackage{microtype}      
\usepackage{xcolor}

\hypersetup{
    colorlinks=true,          
    linkcolor=purple, 
    citecolor=teal,    
    filecolor=magenta,        
    urlcolor=cyan        
}

\usepackage[strict]{changepage}
\usepackage{framed}
\definecolor{customshade}{rgb}{0.941, 0.937, 0.996}
\definecolor{darkblue}{rgb}{0.14,0.22,0.62}
\newenvironment{kkboxline}{%
  \MakeFramed{\advance\hsize-\width\FrameRestore\vspace{-10pt}}%
  \noindent\hspace{-4.55pt}
  \begin{adjustwidth}{}{7pt}%
}
{%
  \end{adjustwidth}\endMakeFramed%
}

\title{\sys: A Multi-Lingual Benchmark for Detecting AI-Generated and Paraphrased Source Code from Production-Level LLMs}

\author{%
  Hanxi Guo \\
  Purdue University\\
  \texttt{guo778@purdue.edu} \\
  \And
 Siyuan Cheng \\
 Purdue University\\
 \texttt{cheng535@purdue.edu}\\
 \And
 Kaiyuan Zhang \\
 Purdue University\\
 \texttt{zhan4057@purdue.edu}\\
 \And
 Guangyu Shen \\
 Purdue University\\
 \texttt{shen447@purdue.edu}\\
 \And
 Xiangyu Zhang \\
 Purdue University\\
 \texttt{xyzhang@purdue.edu}
}

\begin{document}

\maketitle

\begin{abstract}
Large language models (LLMs) have become integral to modern software development, producing vast amounts of AI-generated source code. While these models boost programming productivity, their misuse introduces critical risks, including code plagiarism, license violations, and the propagation of insecure programs. As a result, robust detection of AI-generated code is essential. To support the development of such detectors, a comprehensive benchmark that reflects real-world conditions is crucial. However, existing benchmarks fall short---most cover only a limited set of programming languages and rely on less capable generative models.
In this paper, we present \sys, a comprehensive benchmark that addresses these limitations through three major advancements: (1) it spans ten widely used programming languages, (2) includes both original and paraphrased code samples, and (3) incorporates outputs from ten state-of-the-art production-level LLMs, including both reasoning and non-reasoning models from six major providers. Using \sys, we evaluate ten representative detectors across four methodological paradigms under four realistic evaluation configurations, reporting results using three complementary metrics.
Our analysis reveals nine key findings that uncover the strengths and weaknesses of current detectors, and identify critical challenges for future work. We believe \sys offers a rigorous and practical testbed to advance the development of robust and generalizable AI-generated code detectors.
\end{abstract}

\section{Introduction}
Large Language Models (LLMs) are rapidly evolving and demonstrating increasing capabilities in coding, fundamentally transforming the software development ecosystem. Recent LLMs such as ChatGPT~\cite{chatgpt} and Claude~\cite{claude37} exhibit remarkable code generation performance, producing high-quality outputs in response to concise natural language prompts. The emergence of reasoning-capable models like DeepSeek-R1~\cite{deepseek-r1} has further accelerated LLM adoption among developers. According to Stack Overflow's industry report~\cite{stackoverflow}, 82.1\% of the 65{,}000 surveyed developers report using ChatGPT~\cite{chatgpt} during their development workflow. Capitalizing on the strong coding abilities of LLMs, assistant tools such as GitHub Copilot~\cite{copilot} and Cursor~\cite{cursor} have been developed to enhance productivity by helping developers write, modify, and debug code directly within integrated development environments (IDEs). Furthermore, state-of-the-art LLM-based agentic systems such as OpenHands~\cite{openhands} achieve up to a 65.8\% resolved rate on SWE-Bench~\cite{swebench}, demonstrating the effectiveness of LLMs in addressing real-world software engineering tasks. These trends indicate that LLMs and their associated tools are becoming integral to modern software development workflows.

However, the rapid spread of AI-generated code has raised concerns about new vulnerabilities and misuse.  
Systematic benchmarks show that LLM outputs often ship with logic errors and latent security flaws\citep{liu2023your,gao2023makes,toth2024llms,zhang2023well,pearce2022asleep,kaniewski2024vulnerability}.  
Comparative evaluations reveal that AI suggestions can embed at least as many vulnerabilities as human code\citep{khoury2023secure,wang2024your,tambon2025bugs,vaidya2023critical,asare2023github,tihanyi2025secure}.
Furthermore, LLMs are susceptible to manipulation~\cite{kaniewski2024vulnerability}, including poisoning attacks~\cite{yan2024llm,cotroneo2024vulnerabilities,oh2024poisoned} and prompt injections~\cite{mastropaolo2023robustness,zeng2025inducing}, which can induce the generation of targeted vulnerable code.  
At the same time, educators warn of an impending wave of AI-driven plagiarism that evades conventional detectors~\cite{hutson2024rethinking,steponenaite2023plagiarism,khalaf2025does,dehouche2021plagiarism,xiao2022new,sauglam2024automated,khalil2023will}, while legal scholars highlight intellectual-property~\cite{yu2023codeipprompt,li2023protecting,xu2024llms,stalnaker2024developer} and licence-compliance~\cite{xu2024licoeval} risks.  
Robust AI-code detection is therefore critical for secure software supply chains, responsible academic practice, and licence compliance.  

To address the challenges of AI-generated code identification, various detection methods have been proposed, leveraging statistical features of code~\cite{whodunit}, the capabilities of language models~\cite{xu2024detecting,detectcodegpt,gpt4code,ye2025uncovering,xu2025distinguishing,gptsniffer-journal,gptsniffer-arxiv}, and code embedding models~\cite{suh2025howfar,codexembed}. However, evaluations based on existing benchmarks and datasets~\cite{suh2025howfar,pan2024assessing,aigcodeset,magecode,codetm4,llmgcode} often fall short in three key aspects. First, they typically cover only a narrow set of programming languages---primarily C++ and Python---while neglecting other widely used languages such as Go and HTML, resulting in limited language diversity compared to real-world software development. Second, most benchmarks rely on open-source LLMs with relatively small model sizes and lower generation quality, or include only a small number of commercial models, leaving a gap between benchmark conditions and real-world usage. Third, most existing datasets lack practical adversarial scenarios, such as paraphrasing~\cite{krishna2024paraphrasing,sadasivan2023paraphrase}, which are common in practice and essential for evaluating the robustness of detection systems.
Thus, a rigorous benchmark that captures real-world language diversity, modern commercial models, and adversarial scenarios is indispensable for driving meaningful progress in this emerging field.

We introduce \sys, a comprehensive benchmark for evaluating AI-generated code detectors under realistic and adversarial conditions, to solve the three major limitations identified in prior benchmark work. \sys is constructed from real-world human-written code and enriched with both AI-generated and paraphrased variants produced by a diverse set of state-of-the-art reasoning and non-reasoning LLMs from six major commercial service providers. The paraphrasing techniques are domain-specific and tailored to source code, enabling rigorous evaluation of detector generalization and robustness.

Our key contributions are as follows:
\begin{itemize}
    \item We present a large-scale, multilingual benchmark for AI-generated code detection, spanning 10 widely used programming languages. The dataset comprises approximately 210{,}000 samples, including 10,000 human-written code files sourced from GitHub~\cite{github-code-clean}, as well as AI-generated and paraphrased counterparts produced by 10 production-level LLMs.
    \item We design four progressively challenging evaluation configurations with three complementary performance metrics to facilitate rigorous and realistic assessment of detector effectiveness under various real-world scenarios.
    \item We conduct a comprehensive evaluation of 10 representative detectors across four methodological paradigms using \sys, providing insights into their accuracy, robustness, and generalization across program languages, models, and adversarial settings.
\end{itemize}

\section{Background and Related Work}
\subsection{Taxonomy of AI-Generated Code Detection Methods}
Detecting AI-generated content has been a long-standing challenge in both the natural language~\cite{turingbench,gltr,akram2023empirical,ghosal2023survey} and computer vision domains~\cite{rossler2018faceforensics,guera2018deepfake,zhu2023genimage,dolhansky2020deepfake,zi2020wilddeepfake}, predating even the emergence of large language models (LLMs)~\cite{vaswani2017attention,achiam2023gpt} and diffusion-based generative models~\cite{sohl2015deep,ho2020denoising}. In contrast, detecting AI-generated source code is a relatively new research direction, emerging primarily in the last two years due to the rapid advancements in the coding capabilities of LLMs~\cite{chatgpt,claude37}.

Inspired by traditional statistical-based methods used for AI-generated text detection~\cite{ramos2003using,ippolito-etal-2020-automatic}, early approaches for code focus on analyzing surface-level statistical features. For example, Whodunit~\cite{whodunit} extracts stylometric and complexity-based features from both raw source code and its abstract syntax tree (AST). However, these methods often struggle to distinguish code generated by modern, high-performing LLMs~\cite{chatgpt,claude37,deepseek-r1,gemini-2.0-pro}, which can mimic human coding styles more closely.

To improve detection effectiveness, recent research has explored more advanced techniques---often leveraging large language models (LLMs) or code embedding models---which can be broadly categorized into the following four methodological paradigms:

\parab{Zero-shot Detector.} 
This category of detectors assigns detection confidence scores based on token-level statistics derived from pretrained LLMs, without requiring task-specific fine-tuning. For example, LogRank~\cite{gltr} and Entropy~\cite{entropy} rely on average next-token log-rank and entropy, respectively, to quantify AI-generated token distributions. DetectGPT~\cite{detectgpt} evaluates the divergence between original and perturbed text using a scoring model, which is a strategy extended in code-specific settings by DetectCodeGPT~\cite{detectcodegpt}, GPT4Code~\cite{gpt4code}, and AIGC Detector~\cite{xu2024detecting}, each employing tailored perturbation schemes for code. CR~\cite{ye2025uncovering} instead measures divergence between original and LLM-rewritten code samples. Binoculars~\cite{binoculars} introduces a model-comparison approach, using cross-perplexity between instruction-tuned and non-instruction-tuned LLMs as a detection signal.

\parab{Embedding-based Detector.} 
Embedding-based detectors~\cite{khoury2023secure} utilize pretrained code embedding models, such as CodeT5+ Embedding~\cite{wang2023codet5+} and CodeXEmbed~\cite{codexembed}, to extract high-level semantic representations from either raw source code or abstract syntax trees (ASTs). These embeddings are then fed into lightweight classifiers, \eg, MLP~\cite{rosenblatt1958perceptron}, to perform binary classification between human-written and AI-generated code.

\parab{Fine-tuning-based Detector.} 
This class of detectors fine-tunes transformer-based models to directly capture discriminative patterns between human-written and AI-generated code. For example, GPTSniffer~\cite{gptsniffer-arxiv,gptsniffer-journal} fine-tunes CodeBERT~\cite{codebert} on labeled code samples to perform binary classification. Other approaches~\cite{suh2025howfar} explore different backbone architectures, such as CodeT5+~\cite{wang2023codet5+} and RoBERTa~\cite{roberta}, to enhance detection performance across varied programming languages and generative models.

\parab{Pretrained LLM with Downstream Detector.} 
Unlike zero-shot methods, detectors in this category extract rich semantic representations or statistical signals from pretrained LLMs and train downstream classifiers on these features. For instance, MageCode~\cite{magecode} uses statistical features derived from the hidden state of the classification token in a pretrained CodeT5+~\cite{wang2023codet5+} to train a two-layer linear classifier. Some detectors originally developed for text, such as Raidar~\cite{raidar}, could be extended to code by comparing metrics between original and LLM-rewritten samples, followed by an XGBoost~\cite{xgboost} classifier. BiScope~\cite{biscope} applies a novel bi-directional cross-entropy analysis using pretrained LLMs and feeds the resulting features into a Random Forest~\cite{randomforest} classifier.

\subsection{Existing AI-generated Code Datasets and Benchmarks}
\begin{table}[t]
\centering
\caption{Comparison between existing AI-generated code benchmarks and our \sys.  
\textbf{Gran.} = granularity (\emph{Func}: function/snippet, \emph{Doc}: whole file).  
IID = in–distribution; OOD = out-of-distribution.  
Baseline categories: \textbf{Z} (zero-shot detector), \textbf{E} (embedding-based detector),  
\textbf{F} (fine-tuning-based detector), \textbf{P} (pre-trained LLM + downstream detector).  
Columns “Open LLMs’’ and “Comm.\ LLMs’’ show whether the dataset includes \emph{any} open-source or commercial generators.}
\resizebox{\linewidth}{!}{
\footnotesize
\begin{tabular}{lcccccccccccc}
\toprule
\textbf{Dataset $\downarrow$} \textbf{Stat.$\rightarrow$} &
\makecell{\textbf{\#Lang}} &
\textbf{Gran.} &
\makecell{\textbf{IID}} &
\makecell{\textbf{OOD}} &
\makecell{\textbf{\#Open}\\\textbf{LLMs}} &
\makecell{\textbf{\#Comm.}\\\textbf{LLMs}} &
\makecell{\textbf{Reasoning}\\\textbf{Model}} &
\makecell{\textbf{\#Human}\\\textbf{Code}} &
\makecell{\textbf{\#AI}\\\textbf{Code}} &
\makecell{\textbf{Adv.}\\\textbf{Test}} &
\makecell{\textbf{Quality}\\\textbf{Check}} &
\makecell{\textbf{Baseline}\\\textbf{\#/Cat.}} \\
\midrule
Suh \etal{}\cite{suh2025howfar} & 3 & Func & \cmark & \cmark & 1 & 3 & \xmark & $\sim3.7k$ & $\sim29.5k$ & \xmark & \xmark & 8 / Z,E,F \\
Pan \etal{}\cite{pan2024assessing} & 1 & Func & \cmark & \cmark & 0 & 1 & \xmark & $\sim5k$ & $\sim71k$ & \cmark & \xmark & 5 / Z \\
AIGCodeSet\cite{aigcodeset} & 1 & Func & \cmark & \xmark & 2 & 1 & \xmark & $\sim4.8k$ & $\sim2.9k$ & \xmark & \cmark & 3 / E,F \\
MAGECODE\cite{magecode} & 3 & Doc & \cmark & \xmark & 0 & 3 & \xmark & $\sim81k$ & $\sim45k$ & \xmark & \cmark & 8 / Z \\
CoDet-M4\cite{codetm4} & 3 & Func & \cmark & \cmark & 4 & 1 & \xmark & $\sim252k$ & $\sim246k$ & \cmark & \cmark & 6 / F,P \\
LLMGCode\cite{llmgcode} & 8 & Doc & \cmark & \xmark & 1 & 3 & \xmark & $<1k$ & $2k$ & \xmark & \xmark & 10 / Z,F,P \\
\midrule
\textbf{\sys{} \textit{(Ours)}} & \textbf{10} & \textbf{Doc} & \textbf{\cmark} & \textbf{\cmark} & \textbf{4} & \textbf{6} & \textbf{\cmark} & \boldmath{$10k$} & $\sim$\boldmath{\textbf{$200k$}} & \textbf{\cmark} & \textbf{\cmark} & \textbf{10 / Z,E,F,P} \\
\bottomrule
\end{tabular}}
\vspace{-0.12in}
\label{tab:benchmark-comparison}
\end{table}
Prior studies~\cite{suh2025howfar,pan2024assessing,aigcodeset,magecode,codetm4,llmgcode} has laid important groundwork for building benchmarks to evaluate AI-generated code detectors. As shown in \autoref{tab:benchmark-comparison}, several benchmarks introduce valuable contributions: for instance, Suh~\etal~\cite{suh2025howfar} propose a large-scale function-level dataset spanning three programming languages. Pan~\etal~\cite{pan2024assessing} and CoDet-M4~\cite{codetm4} incorporate adversarial perturbations into AI-generated code to test robustness. AIGCodeSet~\cite{aigcodeset} and MAGECODE~\cite{magecode} employ quality checks during code generation. LLMGCode~\cite{llmgcode} expands language coverage to eight programming languages. Collectively, these datasets serve as solid foundations for evaluating AI-generated code detectors.

However, each of these benchmarks has notable limitations. Most cover only a small number of programming languages, rely on open-source or less capable LLMs, and none of them leverage latest reasoning models~\cite{deepseek-r1,o3-mini,gemini-2.0-flash-thinking}. Furthermore, baseline evaluations in these benchmarks do not comprehensively include all four major categories of detection methods, and only two out of the six existing benchmarks include adversarial testing, which is critical for assessing real-world robustness.

To address these gaps, our proposed benchmark, \sys, includes: (1) code samples across 10 widely used programming languages; (2) outputs from 10 state-of-the-art production-level LLMs, including three reasoning models; (3) both out-of-distribution and adversarial evaluation settings; and (4) baselines covering all four methodological categories of AI-generated code detection.
\section{\sys Framework}

\begin{figure}[t]
    \centering
    \includegraphics[width=1\linewidth]{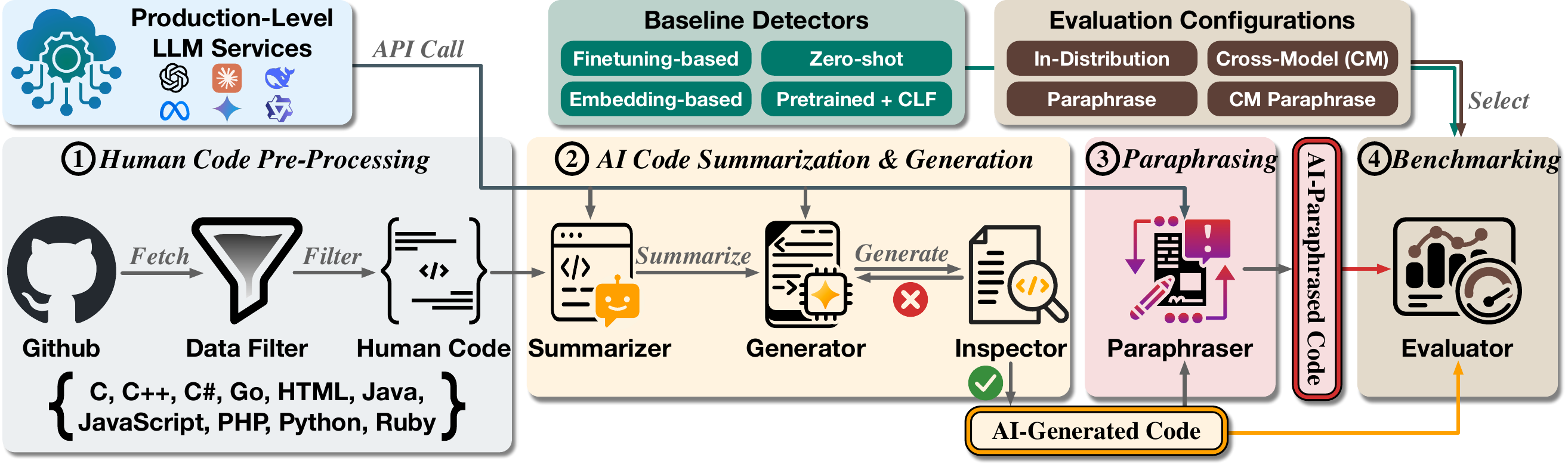}
    \caption{Overview of the \sys framework. We collect and preprocess human-written code from GitHub, then leverage 10 state-of-the-art LLMs to summarize, generate, and paraphrase code with quality inspection. Finally, \sys evaluates 10 baseline AI-generated code detectors across four categories under four configurations, covering a wide range of real-world scenarios.}
    \label{fig:overview}
    \vspace{-0.12in}
\end{figure}

\subsection{Benchmark Construction}
\label{subsection:benchmark construction}

\parab{Human Code Pre-Processing.} To construct a comprehensive benchmark of AI-generated and paraphrased code, we begin by sourcing high-quality human-written code samples from the CodeParrot Github-Code-Clean dataset~\cite{github-code-clean}, a curated subset of the original Github-Code dataset~\cite{github-code}, as shown in \autoref{fig:overview}. This cleaned version filters out overly short snippets, auto-generated files, and samples with excessive alphanumeric characters. The dataset was collected and sanitized in May 2022, prior to the widespread deployment of code LLMs and AI coding agents, ensuring the selected samples are genuinely human-authored. Based on its statistics, we select the ten most commonly used programming languages---C, C++, C\#, Go, HTML, Java, JavaScript, PHP, Python, and Ruby---and randomly extract 1,000 code snippets per language. Additional length-based filtering is applied during the sampling to preserve code diversity while ensuring the code remains within a controlled length scale.

\parab{Production-Level LLMs.} In \sys, we leverage ten production-level LLMs from six leading companies to generate code samples, covering the majority of LLMs commonly used for real-world coding tasks. Among these ten models, four are open-source and three are designed with reasoning capabilities. Specifically, \sys includes GPT-4o-mini~\cite{gpt-4o-mini}, o3-mini~\cite{o3-mini}, Claude-3.5-Haiku~\cite{claude-3.5-haiku}, Gemini-2.0-Flash~\cite{gemini-2.0-flash}, Gemini-2.0-Flash-Thinking-Experimental~\cite{gemini-2.0-flash-thinking}, Gemini-2.0-Pro-Experimental~\cite{gemini-2.0-pro}, DeepSeek-V3~\cite{deepseek-v3}, DeepSeek-R1~\cite{deepseek-r1}, Llama-3.3-70B~\cite{llama-3.3-70b}, and Qwen-2.5-Coder-32B~\cite{qwen-2.5-coder-32b}. We access all ten LLMs via API-based services with default temperatures. For additional details on the LLM configurations and generation settings, please refer to \autoref{appendix:generative model}.

\parab{AI Code Summarization.} To generate high-quality AI-generated code samples while avoiding direct copying of human-written code, \sys adopts a text-to-code generation strategy. As the first step, we produce a comprehensive yet concise summary for each human-written code sample. Since these samples are typically full documents---including library imports, class and structure definitions, and function implementations---we prompt the LLM to extract and summarize key elements such as the \textit{purpose}, \textit{functionality}, \textit{logic overview}, and \textit{key features}, along with the names of relevant \textit{libraries}, \textit{functions}, \textit{classes}, \textit{structures}, and \textit{variables}. \textit{Optional contextual notes} are also included to account for uncommon assumptions or dependencies in the source code. This summary serves as an intermediate representation of the original code, ensuring that the LLM does not access the original human-written implementation during the following code generation step. Full prompts and summary examples are provided in \autoref{appendix:code summarization}.

\parab{AI Code Generation.} Given the summary of each human-written code sample, \sys employs multiple production-level LLMs to generate corresponding AI-written code based on the provided description. To align the structural characteristics of the generated code with the original human-written version, we additionally supply the LLMs with metadata such as the line count and total character length. Due to the inherent uncertainty of LLMs, generated code may occasionally deviate from the desired format or content. To further ensure quality, we implement a rule-based inspector that verifies: (1) consistency with the original human-written code's  line count and character length, and (2) adequate token-level divergence from the original, enforced by requiring a BLEU~\cite{bleu} score below 0.5 to avoid recitation. Regeneration is forced if any check fails, and samples are discarded after multiple failed attempts. Detailed prompts and generation examples are provided in \autoref{appendix:code generation}.

\parab{AI Code Paraphrasing.} Paraphrasing~\cite{krishna2024paraphrasing, sadasivan2023paraphrase} is a widely adopted strategy for evaluating the robustness of AI-generated text detectors under adversarial and real-world conditions. However, in the domain of AI-generated code detection, most existing benchmarks~\cite{suh2025howfar, pan2024assessing, aigcodeset, magecode, codetm4, llmgcode} do not incorporate such adversarial testing. Although some text detection studies~\cite{raidar, biscope} have included paraphrased code in their evaluations, they rely on generic prompts and a limited number of code samples, constraining both the effectiveness and generality of their paraphrasing evaluation on code. In \sys, we introduce a systematic, domain-specific paraphrasing for code, covering six transformation types: \textit{renaming}, \textit{formatting adjustments}, \textit{logic rewriting and replacement}, \textit{expression variation}, \textit{literal transformations}, and \textit{redundancy insertion}. Detailed rules, prompt designs, and representative examples are provided in \autoref{appendix:code paraphrase}.

\subsection{Benchmark Statistics}
\label{sec:benchmark statistics}
\begin{figure}[t]
    \centering
    \includegraphics[width=1\linewidth]{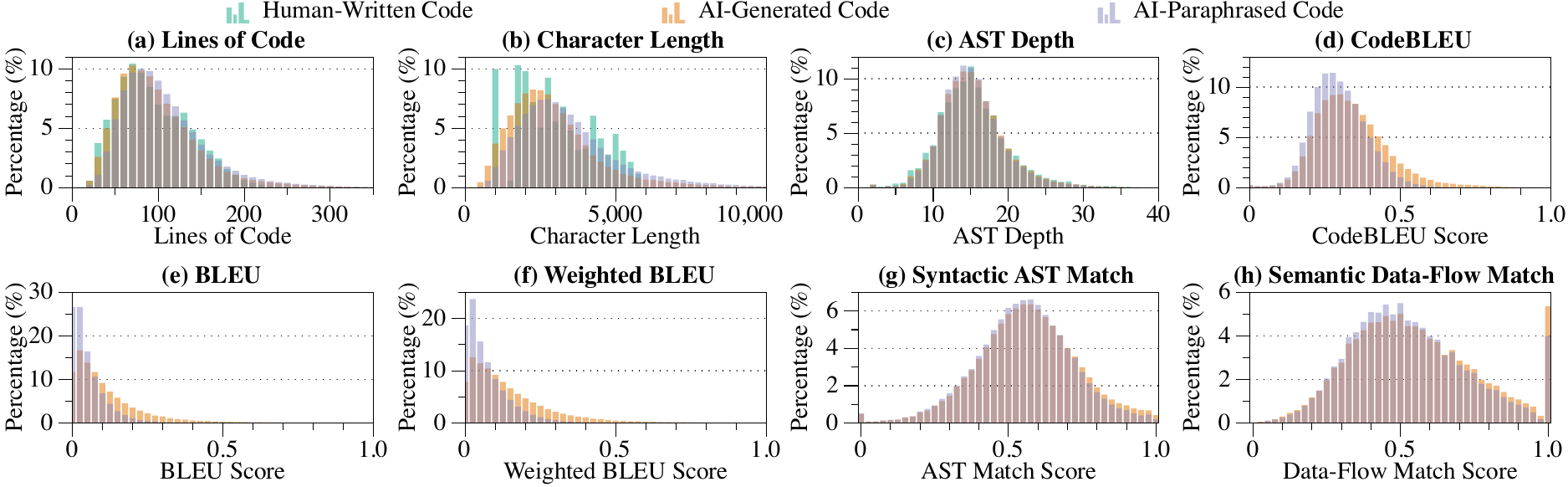}
    \caption{Benchmark statistics of \sys.}
    \label{fig:benchmark-statistics}
    \vspace{-0.12in}
\end{figure}
\sys spans ten programming languages, each containing 1,000 human‑written code samples and 10,000 AI‑generated counterparts. For every language, we obtain 1,000 outputs from each of ten production‑level LLMs, yielding a 1{:}10 mapping between every human sample and its LLM‑generated variants. Within every 1,000‑sample shard (human or AI), we allocate 700 examples for training and 300 for testing. 

We present four structural and semantic metrics of the dataset in \autoref{fig:benchmark-statistics}: lines of code (a), character length (b), AST depth (c), and CodeBLEU~\cite{codebleu} score (d). The first three metrics reflect the overall structural characteristics of the code and show close resemblance between human-written and AI-generated samples. This similarity implies that naive statistical classifiers would struggle to detect AI-generated code using basic code features.

\autoref{fig:benchmark-statistics} (d) reports the CodeBLEU score, a composite metric calculated as:
\begin{equation}
    CodeBLEU = \alpha\cdot BLEU + \beta\cdot BLEU_{weighted} + \gamma\cdot Match_{AST} + \delta\cdot Match_{DF},
\end{equation}
where each component is equally weighted with $\alpha = \beta = \gamma = \delta = 0.25$ by default. The median CodeBLEU score for AI-generated code is approximately 0.3, consistent with prior observations in text-to-code generation~\cite{dong2023codep,espejel2023jacotext,evtikhiev2023out}. Paraphrased code yields slightly lower scores due to deliberate perturbations in both code format and structure.

To further analyze \sys's code quality, we decompose the CodeBLEU score into its four subcomponents in \autoref{fig:benchmark-statistics} (e)–(h). Both AI-generated and AI-paraphrased code show relatively low BLEU~\cite{bleu} and weighted BLEU~\cite{codebleu} scores, indicating limited n-gram overlap with their human counterparts. While the syntactic AST match and semantic data-flow~\cite{dataflow} match scores of AI code exceed 0.5 on average, suggesting that despite token-level divergence, both AI-generated and AI-paraphrased code maintains a fair level of syntactic and semantic consistency with human-written code. More detailed benchmark statistics are presented in \autoref{appendix:data statistics}.

\subsection{Baseline Detectors} \label{sec:baseline_desc}
We select ten state-of-the-art detectors spanning four categories. \textbf{Zero‑shot detectors}: \emph{LogRank}~\cite{gltr}, \emph{Entropy}~\cite{gltr,entropy}, and \emph{Binoculars}~\cite{binoculars}, which rely on token‑rank or entropy-related features without training. \textbf{Embedding‑based detectors}: following existing studies~\cite{suh2025howfar}, we extract representations with the \emph{CodeXEmbed‑2B} model~\cite{codexembed} from either raw source code or its abstract‑syntax tree (AST) and train a lightweight random forest~\cite{randomforest} classifier. \textbf{Fine‑tuned detectors}: we include \emph{GPTSniffer}~\cite{gptsniffer-journal,gptsniffer-arxiv}, a variant built on the latest \emph{CodeT5+} backbone~\cite{wang2023codet5+}, and a \emph{RoBERTa} detector~\cite{roberta}, with each fine‑tuned on our training corpus. \textbf{Pretrained‑LLM with downstream detector}: \emph{Raidar}~\cite{raidar} and \emph{BiScope}~\cite{biscope}, extracting features via rewriting~\cite{raidar} and bi-directional cross entropy~\cite{biscope}. More details of the baseline detectors are presented in \autoref{appendix:baseline}.

\subsection{Evaluation Metrics}
To thoroughly assess the performance of the baseline detectors in different scenarios, we employ three evaluation metrics in our experiments, including the \textit{F1 score}, \textit{TPR@FPR=10\%}, and \textit{TPR@FPR=1\%}.
The \textit{F1 score} balances precision and recall, providing an overall measure of detection accuracy without favoring AI-generated or human-written code samples. For each detector, we first identify the optimal decision threshold and then report its corresponding \textit{F1 score}. The metric \textit{TPR@FPR=10\%} reports the true positive rate (TPR) when the false positive rate (FPR) is limited to 10\%, representing scenarios that can tolerate a moderate number of false alarms. Conversely, \textit{TPR@FPR=1\%} measures the TPR at an FPR of only 1\%, which is essential for applications where even a small fraction of false positives is unacceptable.

\subsection{Evaluation Configurations} \label{sec:eval_config}
In \sys, we include four evaluation configurations to thoroughly assess baseline detectors under various real-world scenarios, including the in-distribution configuration and three out-of-distribution configurations (paraphrase configuration, cross-model configuration, and cross-model paraphrase configuration). We omit the cross language configuration because programming language can be easily identified; thus, detectors can be trained separately for each language.

\parab{In-Distribution Configuration.} This configuration evaluates the in-distribution stability of each detector in multiple LLMs and programming languages. For each language, we pair the human‑written training set with the training samples produced by a single LLM, train the detector on this combined data, and determine the optimal decision threshold. We then test the detector on the human‑written test set together with the test samples generated by the same LLM. 

\parab{Paraphrase Configuration.} This setting evaluates each detector’s out‑of‑distribution performance when the AI‑generated code is adversarially paraphrased. Specifically, we train the detector and select its optimal threshold same as in the in‑distribution configuration, but we test on \emph{paraphrased} code produced by the same LLM that generated the original samples.

\parab{Cross-Model Configuration.} This setting evaluates detector's robustness against unseen LLMs. For each programming language, we train the detector and choose its optimal threshold on a training set consisting of human‑written samples and AI‑generated samples from a \emph{single} LLM. We then test the detector on human test samples paired with AI‑generated samples from all \emph{other} LLMs. The detector’s scores on these unseen‑model test sets are averaged to yield the overall cross‑model result.

\parab{Cross-Model Paraphrase Configuration.} This scenario mirrors real‑world conditions in which code samples are both generated by unseen LLMs and subsequently paraphrased. We adopt the testing procedure of the cross‑model configuration, but pair human test samples with paraphrased test samples produced by the other LLMs. The detector’s average score over all such paraphrased, unseen‑model test sets is reported as the cross‑model paraphrase result.

\section{Evaluation Results and Insights}
\label{section:evaluation}
We conduct an extensive evaluation using \sys in various scenarios and summarize the observations into nine findings. We present representative processed results in the main text and include the full experimental results in \autoref{appendix:results}.

\subsection{Comparison Between Evaluation Configurations and Detectors}

\begin{figure}[t]
    \centering
    \includegraphics[width=1\linewidth]{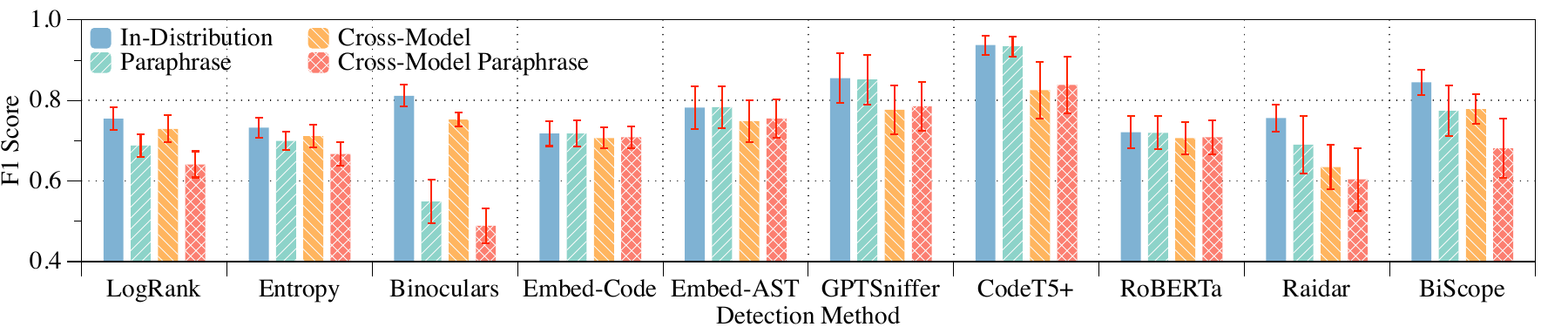}
    \caption{\textbf{Comparison Between Evaluation Configurations and Detectors.} The bar chart presents the average F1 scores of baseline detectors across all the programming languages and LLMs.}
    \label{fig:across-language-llm}
    \vspace{-0.12in}
\end{figure}

We first evaluate the performance of the various detectors under four distinct configurations~\ref{sec:eval_config}.
The results are presented in \autoref{fig:across-language-llm}, where the x-axis lists the detectors and the y-axis represents the F1 score. Each bar corresponds to a specific evaluation configuration.
Notably, to ensure a fair and unbiased comparison, each bar reflects the average F score obtained across ten programming languages and ten LLMs, with error bars indicating one standard deviation.

\begin{kkboxline}
\textbf{Finding 1:} 
\textit{In-distribution testing consistently outperforms all out-of-distribution scenarios.}
\end{kkboxline}

This is intuitive and reasonable given the shared distribution between training and test sets.
Under out-of-distribution settings, cross-model testing yields a larger performance drop than paraphrasing in most cases, since paraphrasing leverages the same LLM and thus incurs a smaller distribution shift than code generation by a different LLM.
However, some corner cases, e.g., LogRank and Binoculars, deviate from this trend. As zero-shot methods, they are particularly sensitive to token-level features, and paraphrasing induces greater token variance than cross-model evaluation.

Furthermore, different detection methods exhibit varying performance.
According to \autoref{sec:baseline_desc}, these methods fall into four categories.

\begin{kkboxline}
\textbf{Finding 2:} 
\textit{Fine-tuning-based methods outperforms other types.}
\end{kkboxline}

Fine-tuned detectors, e.g., GPTSniffer and CodeT5+, lead the pack.
Zero-shot approaches, e.g., LogRank and Entropy, perform poorest, which makes sense given their limited feature extraction when confronted with the complexity of code.
Embedding-based detectors, e.g., Embed-Code and Embed-AST, sit in the middle but impressively maintain stable accuracy even under out-of-distribution evaluation, thanks to their reliance on code representations that generalize across LLMs.
Pretrained LLMs paired with downstream classifiers, e.g., Raidar and BiScope, match embedding methods in-distribution but suffer a larger drop on out-of-distribution tests, reflecting subtle shifts in the features they extract across different models and paraphrased inputs.

\begin{kkboxline}
\textbf{Finding 3:} 
\textit{Fine-tuning approaches using backbone LLMs pre-trained on larger code corpora achieve superior performance.}
\end{kkboxline}

Performance varies across fine-tuning methods. For example, CodeT5+ slightly outperforms GPTSniffer, and both surpass RoBERTa. This gap reflects their pre-training corpora: GPTSniffer’s CodeBERT backbone is trained on six programming languages, whereas CodeT5+’s backbone covers nine. In contrast, RoBERTa is pretrained solely on natural-language text. Consequently, backbones exposed to more and broader code samples exhibit superior coding proficiency, and hence better detection capability.

\begin{kkboxline}
\textbf{Finding 4:} 
\textit{Fine-tuning–based detectors are prone to overfitting.}
\end{kkboxline}
We also observe that fine-tuning–based methods (e.g., GPTSniffer and CodeT5+) exhibit a larger performance drop from in-distribution to cross-model evaluations than other approaches. This is likely due to their overfitting tendencies and should be taken into account in real-world deployments.

\begin{kkboxline}
\textbf{Finding 5:} 
\textit{ASTs provide a superior feature representation compared to raw source code.}
\end{kkboxline}
Two embedding–based detectors demonstrate comparable performance, with Embed-AST marginally outperforming Embed-Code. This suggests that AST-based embeddings capture the program’s syntactic hierarchy and semantic relationships, e.g., control flow and data dependencies, more effectively than raw code tokens, making them more robust to superficial variations like naming or formatting.

\subsection{Comparison Between Different Programming Languages}
\begin{figure}[!htbp]
    \centering
    \includegraphics[width=1\linewidth]{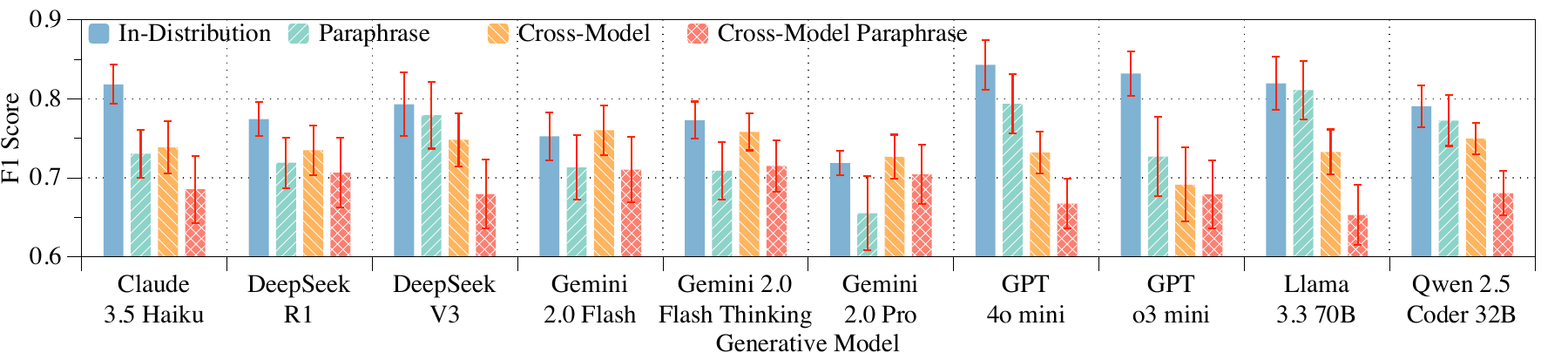}
    \caption{\textbf{Comparison Between Different LLMs.} The bar chart shows the average F1 scores of baseline detectors on different LLMs across programming languages.}
    \label{fig:across-language-detector}
    \vspace{-0.12in}
\end{figure}

We evaluate detection performance across ten programming languages using \sys. The results are shown in \autoref{fig:across-language-detector}, where the x-axis lists the languages and the y-axis denotes the F1 score.
To minimize bias, each bar aggregates results from experiments with all ten LLMs and ten detectors. Its height indicates the average F1 score, and the error bars represent one standard deviation.

\begin{figure}[t]
    \centering
    \includegraphics[width=1\linewidth]{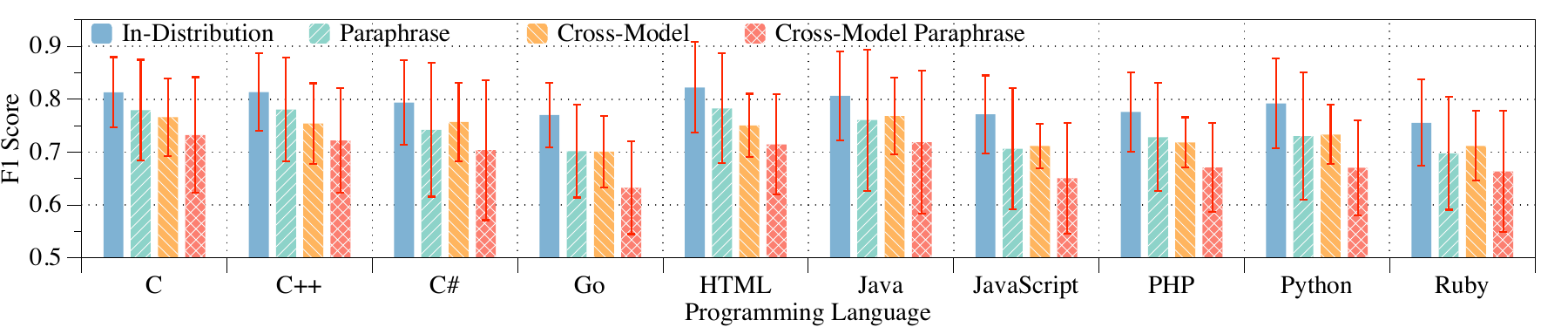}
    \caption{\textbf{Comparison Between Different Programming Languages.} The bar chart presents the average F1 scores of baseline detectors on different programming languages across LLMs.}
    \label{fig:across-detector-llm}
    \vspace{-0.12in}
\end{figure}

\begin{kkboxline}
\textbf{Finding 6:} 
\textit{Detection is Consistent across Programming Languages, with Common Languages Performing Slightly Better.}
\end{kkboxline}

We observe only slight performance differences among languages, with similar patterns across evaluation configurations.
Notably, less common languages exhibit marginally lower performance. For example, C++ achieves higher F1 scores than Go or Ruby. This discrepancy arises because several detection methods, e.g., Biscope~\cite{biscope} and Raidar~\cite{raidar}, rely on pre-trained LLMs for feature extraction. These models are pre-trained on large online corpora containing more examples of common languages (e.g., C++) than atypical ones (e.g., Go), resulting in stronger representations for the former.
Hence detection performances are better detection on those common languages.

\subsection{Comparison Between Different LLMs}

We evaluate the detection performance of code generated by different LLMs, with results shown in \autoref{fig:across-language-detector}. The x-axis represents the generative models, while the y-axis indicates the F1 score. Each bar color corresponds to one of four evaluation settings.

\begin{kkboxline}
\textbf{Finding 7:} 
\textit{Detection performance is generally similar across LLMs, with GPT and Llama showing slightly higher scores.}
\end{kkboxline}

Among all models, GPT-4o mini achieves the highest F1 scores, particularly under the In-Distribution and Paraphrase settings, suggesting that its code style is more consistent or distinctive, making detection easier. Claude 3.5 Haiku and Llama 3.3 70B also demonstrate strong performance, especially under In-Distribution, likely due to their more recognizable or less variable code patterns.
In contrast, Cross-Model Paraphrase consistently yields the lowest F1 scores (around 0.65–0.7), highlighting it as the most challenging scenario for detection. Models such as Gemini 2.0 Pro and Qwen 2.5 Coder 32B exhibit lower detectability across settings, especially under paraphrased or cross-model conditions, indicating that their outputs may be more diverse or stylistically more similar to human's, thereby reducing their distinctiveness.

\begin{kkboxline}
\textbf{Finding 8:} 
\textit{Reasoning models exhibit a larger performance drop after paraphrasing.}
\end{kkboxline}

We observe that for non-reasoning models (DeepSeek V3, GPT4o mini, Llama 3.3 70B, and Qwen 2.5 Coder 32B), paraphrasing has minimal impact on performance. In contrast, reasoning models (e.g., GPT o3 mini) suffer a more pronounced decline. This likely stems from their stronger comprehension abilities: they better interpret paraphrased inputs and adjust outputs to match human-style reasoning, making any deviations more evident after paraphrasing.

\subsection{Comparison Between Different Evaluation Metrics}

In previous experiments, we mainly use F1 score, which is a threshold‐dependent measure that balances precision and recall, but F1 can be misleading in real‐world detection tasks. As it gives equal weight to false positives and false negatives and depends on a single decision threshold, it often fails to reflect performance in imbalanced settings or under strict false‐alarm constraints.
By contrast, reporting the true positive rate at low false‐positive rates directly measures how many genuine positives the model catches when false alarms must be kept to a minimum~\cite{carlini2022membership}.
Therefore, we introduce two additional metrics, i.e., TPR@FPR=10\% and 1\%, to better assess detector practicality.

\begin{kkboxline}
\textbf{Finding 9:} 
\textit{There is a significant gap between laboratory evaluations and practical use.}
\end{kkboxline}

Results in \autoref{appendix:tpr} indicates that despite decent F1 scores, all detectors suffer a dramatic drop in true‐positive rate once the false-positive rate is constrained, showing that they fail to catch enough positives under realistic, low-alarm requirements and are therefore impractical.
\section{Conclusion}
In this paper, we introduce \sys, a comprehensive and large-scale benchmark for AI-generated code detection, consisting of 10 widely used programming languages and approximately 210,000 samples in total. The dataset includes human-written code, as well as AI-generated and paraphrased variants created using 10 state-of-the-art production-level LLMs, including three recent reasoning models, with quality control to ensure reliability. We evaluate 10 representative detectors spanning four methodological categories and provide extensive analysis from multiple perspectives, revealing key strengths and limitations of each approach. We believe the breadth and depth of \sys offer a strong foundation for advancing the development of more robust and generalizable detectors.

{\small
\bibliography{reference}

\begin{thebibliography}{10}

\bibitem{achiam2023gpt}
Josh Achiam, Steven Adler, Sandhini Agarwal, Lama Ahmad, Ilge Akkaya, Florencia~Leoni Aleman, Diogo Almeida, Janko Altenschmidt, Sam Altman, Shyamal Anadkat, et~al.
\newblock Gpt-4 technical report.
\newblock {\em arXiv preprint arXiv:2303.08774}, 2023.

\bibitem{akram2023empirical}
Arslan Akram.
\newblock An empirical study of ai generated text detection tools.
\newblock {\em arXiv preprint arXiv:2310.01423}, 2023.

\bibitem{claude-3.5-haiku}
Anthropic.
\newblock Introducing computer use, a new claude 3.5 sonnet, and claude 3.5 haiku, 2024.

\bibitem{claude37}
{Anthropic}.
\newblock {Claude 3.7 Sonnet and Claude Code}, 2025.

\bibitem{asare2023github}
Owura Asare, Meiyappan Nagappan, and Nirmal Asokan.
\newblock Is github’s copilot as bad as humans at introducing vulnerabilities in code?
\newblock {\em Empirical Software Engineering}, 28(6):129, 2023.

\bibitem{randomforest}
Leo Breiman.
\newblock Random forests.
\newblock {\em Machine learning}, 45:5--32, 2001.

\bibitem{carlini2022membership}
Nicholas Carlini, Steve Chien, Milad Nasr, Shuang Song, Andreas Terzis, and Florian Tramer.
\newblock Membership inference attacks from first principles.
\newblock In {\em 2022 IEEE symposium on security and privacy (SP)}, pages 1897--1914. IEEE, 2022.

\bibitem{xgboost}
Tianqi Chen, Tong He, Michael Benesty, Vadim Khotilovich, Yuan Tang, Hyunsu Cho, Kailong Chen, Rory Mitchell, Ignacio Cano, Tianyi Zhou, Mu~Li, Junyuan Xie, Min Lin, Yifeng Geng, Yutian Li, Jiaming Yuan, and David Cortes.
\newblock {\em xgboost: Extreme Gradient Boosting}, 2025.
\newblock R package version 3.0.1.1.

\bibitem{github-code-clean}
CodeParrot.
\newblock Github code clean dataset, 2022.

\bibitem{github-code}
CodeParrot.
\newblock Github code dataset, 2022.

\bibitem{cotroneo2024vulnerabilities}
Domenico Cotroneo, Cristina Improta, Pietro Liguori, and Roberto Natella.
\newblock Vulnerabilities in ai code generators: Exploring targeted data poisoning attacks.
\newblock In {\em IEEE/ACM International Conference on Program Comprehension (ICPC)}, pages 280--292, 2024.

\bibitem{cursor}
{Cursor}.
\newblock {Cursor: The AI Code Editor}, 2023.

\bibitem{dehouche2021plagiarism}
Nassim Dehouche.
\newblock Plagiarism in the age of massive generative pre-trained transformers (gpt-3).
\newblock {\em Ethics in Science and Environmental Politics}, 21:17--23, 2021.

\bibitem{aigcodeset}
Basak Demirok and Mucahid Kutlu.
\newblock Aigcodeset: A new annotated dataset for ai generated code detection.
\newblock {\em arXiv preprint arXiv:2412.16594}, 2024.

\bibitem{dolhansky2020deepfake}
Brian Dolhansky, Joanna Bitton, Ben Pflaum, Jikuo Lu, Russ Howes, Menglin Wang, and Cristian~Canton Ferrer.
\newblock The deepfake detection challenge (dfdc) dataset.
\newblock {\em arXiv preprint arXiv:2006.07397}, 2020.

\bibitem{dong2023codep}
Yihong Dong, Ge~Li, and Zhi Jin.
\newblock Codep: grammatical seq2seq model for general-purpose code generation.
\newblock In {\em ACM SIGSOFT International Symposium on Software Testing and Analysis (ISSTA)}, pages 188--198, 2023.

\bibitem{espejel2023jacotext}
Jessica~L{\'o}pez Espejel, Mahaman Sanoussi~Yahaya Alassan, Walid Dahhane, and El~Hassane Ettifouri.
\newblock Jacotext: a pretrained model for java code-text generation.
\newblock {\em arXiv preprint arXiv:2303.12869}, 2023.

\bibitem{evtikhiev2023out}
Mikhail Evtikhiev, Egor Bogomolov, Yaroslav Sokolov, and Timofey Bryksin.
\newblock Out of the bleu: how should we assess quality of the code generation models?
\newblock {\em Journal of Systems and Software}, 203:111741, 2023.

\bibitem{codebert}
Zhangyin Feng, Daya Guo, Duyu Tang, Nan Duan, Xiaocheng Feng, Ming Gong, Linjun Shou, Bing Qin, Ting Liu, Daxin Jiang, and Ming Zhou.
\newblock Codebert: A pre-trained model for programming and natural languages.
\newblock {\em https://arxiv.org/abs/2002.08155}, 2020.

\bibitem{copilot}
{Friedman, Nat}.
\newblock {Introducing GitHub Copilot: your AI pair programmer}, 2022.

\bibitem{gao2023makes}
Shuzheng Gao, Xin-Cheng Wen, Cuiyun Gao, Wenxuan Wang, Hongyu Zhang, and Michael~R Lyu.
\newblock What makes good in-context demonstrations for code intelligence tasks with llms?
\newblock In {\em IEEE/ACM International Conference on Automated Software Engineering (ASE)}, pages 761--773, 2023.

\bibitem{gltr}
Sebastian Gehrmann, Hendrik Strobelt, and Alexander~M Rush.
\newblock Gltr: Statistical detection and visualization of generated text.
\newblock In {\em Annual Meeting of the Association for Computational Linguistics (ACL)}, 2019.

\bibitem{ghosal2023survey}
Soumya~Suvra Ghosal, Souradip Chakraborty, Jonas Geiping, Furong Huang, Dinesh Manocha, and Amrit Bedi.
\newblock A survey on the possibilities \& impossibilities of ai-generated text detection.
\newblock {\em Transactions on Machine Learning Research (TMLR)}, 2023.

\bibitem{guera2018deepfake}
David G{\"u}era and Edward~J Delp.
\newblock Deepfake video detection using recurrent neural networks.
\newblock In {\em IEEE International Conference on Advanced Video and Signal Based Surveillance (AVSS)}, pages 1--6, 2018.

\bibitem{dataflow}
Daya Guo, Shuo Ren, Shuai Lu, Zhangyin Feng, Duyu Tang, LIU Shujie, Long Zhou, Nan Duan, Alexey Svyatkovskiy, Shengyu Fu, et~al.
\newblock Graphcodebert: Pre-training code representations with data flow.
\newblock In {\em International Conference on Learning Representations (ICLR)}, 2021.

\bibitem{deepseek-r1}
Daya Guo, Dejian Yang, Haowei Zhang, Junxiao Song, Ruoyu Zhang, Runxin Xu, Qihao Zhu, Shirong Ma, Peiyi Wang, Xiao Bi, et~al.
\newblock Deepseek-r1: Incentivizing reasoning capability in llms via reinforcement learning.
\newblock {\em arXiv preprint arXiv:2501.12948}, 2025.

\bibitem{biscope}
Hanxi Guo, Siyuan Cheng, Xiaolong Jin, Zhuo Zhang, Kaiyuan Zhang, Guanhong Tao, Guangyu Shen, and Xiangyu Zhang.
\newblock Biscope: Ai-generated text detection by checking memorization of preceding tokens.
\newblock {\em Advances in Neural Information Processing Systems (NeurIPS)}, 37:104065--104090, 2024.

\bibitem{binoculars}
Abhimanyu Hans, Avi Schwarzschild, Valeriia Cherepanova, Hamid Kazemi, Aniruddha Saha, Micah Goldblum, Jonas Geiping, and Tom Goldstein.
\newblock Spotting llms with binoculars: Zero-shot detection of machine-generated text.
\newblock In {\em International Conference on Machine Learning (ICML)}, 2024.

\bibitem{ho2020denoising}
Jonathan Ho, Ajay Jain, and Pieter Abbeel.
\newblock Denoising diffusion probabilistic models.
\newblock {\em Advances in Neural Information Processing Systems (NeurIPS)}, 33:6840--6851, 2020.

\bibitem{qwen-2.5-coder-32b}
Binyuan Hui, Jian Yang, Zeyu Cui, Jiaxi Yang, Dayiheng Liu, Lei Zhang, Tianyu Liu, Jiajun Zhang, Bowen Yu, Keming Lu, et~al.
\newblock Qwen2. 5-coder technical report.
\newblock {\em arXiv preprint arXiv:2409.12186}, 2024.

\bibitem{hutson2024rethinking}
James Hutson.
\newblock Rethinking plagiarism in the era of generative ai.
\newblock {\em Journal of Intelligent Communication}, 3(2):20--31, 2024.

\bibitem{whodunit}
Oseremen~Joy Idialu, Noble~Saji Mathews, Rungroj Maipradit, Joanne~M Atlee, and Mei Nagappan.
\newblock Whodunit: Classifying code as human authored or gpt-4 generated-a case study on codechef problems.
\newblock In {\em International Conference on Mining Software Repositories (MSR)}, pages 394--406, 2024.

\bibitem{ippolito-etal-2020-automatic}
Daphne Ippolito, Daniel Duckworth, Chris Callison-Burch, and Douglas Eck.
\newblock Automatic detection of generated text is easiest when humans are fooled.
\newblock In {\em Annual Meeting of the Association for Computational Linguistics (ACL)}, pages 1808--1822, 2020.

\bibitem{swebench}
Carlos~E Jimenez, John Yang, Alexander Wettig, Shunyu Yao, Kexin Pei, Ofir Press, and Karthik~R Narasimhan.
\newblock Swe-bench: Can language models resolve real-world github issues?
\newblock In {\em International Conference on Learning Representations (ICLR)}, 2024.

\bibitem{gemini-2.0-flash-thinking}
Patrick Kane.
\newblock Access the latest 2.0 experimental models in the gemini app., 2025.

\bibitem{kaniewski2024vulnerability}
Sabrina Kaniewski, Dieter Holstein, Fabian Schmidt, and Tobias Heer.
\newblock Vulnerability handling of ai-generated code-existing solutions and open challenges.
\newblock In {\em Conference on AI, Science, Engineering, and Technology (AIxSET)}, pages 145--148, 2024.

\bibitem{gemini-2.0-pro}
Koray Kavukcuoglu.
\newblock Gemini 2.0 is now available to everyone, 2025.

\bibitem{khalaf2025does}
Mustafa~Ali Khalaf.
\newblock Does attitude towards plagiarism predict aigiarism using chatgpt?
\newblock {\em AI and Ethics}, 5(1):677--688, 2025.

\bibitem{khalil2023will}
Mohammad Khalil and Erkan Er.
\newblock Will chatgpt g et you caught? rethinking of plagiarism detection.
\newblock In {\em International Conference on Human-Computer Interaction}, pages 475--487, 2023.

\bibitem{khoury2023secure}
Rapha{\"e}l Khoury, Anderson~R Avila, Jacob Brunelle, and Baba~Mamadou Camara.
\newblock How secure is code generated by chatgpt?
\newblock In {\em IEEE international conference on systems, man, and cybernetics (SMC)}, pages 2445--2451. IEEE, 2023.

\bibitem{krishna2024paraphrasing}
Kalpesh Krishna, Yixiao Song, Marzena Karpinska, John Wieting, and Mohit Iyyer.
\newblock Paraphrasing evades detectors of ai-generated text, but retrieval is an effective defense.
\newblock {\em Advances in Neural Information Processing Systems (NeurIPS)}, 2023.

\bibitem{entropy}
Thomas Lavergne, Tanguy Urvoy, and Fran{\c{c}}ois Yvon.
\newblock Detecting fake content with relative entropy scoring.
\newblock In {\em Proceedings of the International Conference on Uncovering Plagiarism, Authorship and Social Software Misuse (PAN)}, volume 377, pages 27--31, 2008.

\bibitem{li2023protecting}
Zongjie Li, Chaozheng Wang, Shuai Wang, and Cuiyun Gao.
\newblock Protecting intellectual property of large language model-based code generation apis via watermarks.
\newblock In {\em ACM SIGSAC Conference on Computer and Communications Security (CCS)}, pages 2336--2350, 2023.

\bibitem{deepseek-v3}
Aixin Liu, Bei Feng, Bing Xue, Bingxuan Wang, Bochao Wu, Chengda Lu, Chenggang Zhao, Chengqi Deng, Chenyu Zhang, Chong Ruan, et~al.
\newblock Deepseek-v3 technical report.
\newblock {\em arXiv preprint arXiv:2412.19437}, 2024.

\bibitem{liu2023your}
Jiawei Liu, Chunqiu~Steven Xia, Yuyao Wang, and Lingming Zhang.
\newblock Is your code generated by chatgpt really correct? rigorous evaluation of large language models for code generation.
\newblock {\em Advances in Neural Information Processing Systems (NeurIPS)}, 36:21558--21572, 2023.

\bibitem{codexembed}
Ye~Liu, Rui Meng, Shafiq Joty, Silvio Savarese, Caiming Xiong, Yingbo Zhou, and Semih Yavuz.
\newblock Codexembed: A generalist embedding model family for multiligual and multi-task code retrieval.
\newblock {\em arXiv preprint arXiv:2411.12644}, 2024.

\bibitem{roberta}
Yinhan Liu, Myle Ott, Naman Goyal, Jingfei Du, Mandar Joshi, Danqi Chen, Omer Levy, Mike Lewis, Luke Zettlemoyer, and Veselin Stoyanov.
\newblock Roberta: A robustly optimized bert pretraining approach.
\newblock {\em arXiv preprint arXiv:1907.11692}, 2019.

\bibitem{raidar}
Chengzhi Mao, Carl Vondrick, Hao Wang, and Junfeng Yang.
\newblock Raidar: generative ai detection via rewriting.
\newblock In {\em International Conference on Learning Representations (ICLR)}, 2024.

\bibitem{mastropaolo2023robustness}
Antonio Mastropaolo, Luca Pascarella, Emanuela Guglielmi, Matteo Ciniselli, Simone Scalabrino, Rocco Oliveto, and Gabriele Bavota.
\newblock On the robustness of code generation techniques: An empirical study on github copilot.
\newblock In {\em International Conference on Software Engineering (ICSE)}, pages 2149--2160, 2023.

\bibitem{llama-3.3-70b}
Meta.
\newblock Llama 3.3: Model cards \& prompt formats, 2024.

\bibitem{detectgpt}
Eric Mitchell, Yoonho Lee, Alexander Khazatsky, Christopher~D Manning, and Chelsea Finn.
\newblock Detectgpt: Zero-shot machine-generated text detection using probability curvature.
\newblock In {\em International Conference on Machine Learning (ICML)}, pages 24950--24962. PMLR, 2023.

\bibitem{gptsniffer-arxiv}
Phuong~T Nguyen, Juri Di~Rocco, Claudio Di~Sipio, Riccardo Rubei, Davide Di~Ruscio, and Massimiliano Di~Penta.
\newblock Is this snippet written by chatgpt? an empirical study with a codebert-based classifier.
\newblock {\em arXiv preprint arXiv:2307.09381}, 2023.

\bibitem{gptsniffer-journal}
Phuong~T Nguyen, Juri Di~Rocco, Claudio Di~Sipio, Riccardo Rubei, Davide Di~Ruscio, and Massimiliano Di~Penta.
\newblock Gptsniffer: A codebert-based classifier to detect source code written by chatgpt.
\newblock {\em Journal of Systems and Software}, 214:112059, 2024.

\bibitem{oh2024poisoned}
Sanghak Oh, Kiho Lee, Seonhye Park, Doowon Kim, and Hyoungshick Kim.
\newblock Poisoned chatgpt finds work for idle hands: Exploring developers’ coding practices with insecure suggestions from poisoned ai models.
\newblock In {\em IEEE Symposium on Security and Privacy (S\&P)}, pages 1141--1159, 2024.

\bibitem{chatgpt}
{OpenAI}.
\newblock {Introducing ChatGPT}, 2022.

\bibitem{gpt-4o-mini}
OpenAI.
\newblock Gpt-4o mini: advancing cost-efficient intelligence, 2024.

\bibitem{o3-mini}
OpenAI.
\newblock Openai o3-mini: Pushing the frontier of cost-effective reasoning, 2025.

\bibitem{codetm4}
Daniil Orel, Dilshod Azizov, and Preslav Nakov.
\newblock Codet-m4: Detecting machine-generated code in multi-lingual, multi-generator and multi-domain settings.
\newblock {\em arXiv preprint arXiv:2503.13733}, 2025.

\bibitem{pan2024assessing}
Wei~Hung Pan, Ming~Jie Chok, Jonathan Leong~Shan Wong, Yung~Xin Shin, Yeong~Shian Poon, Zhou Yang, Chun~Yong Chong, David Lo, and Mei~Kuan Lim.
\newblock Assessing ai detectors in identifying ai-generated code: Implications for education.
\newblock In {\em International Conference on Software Engineering: Software Engineering Education and Training (ICSE-SEET)}, pages 1--11, 2024.

\bibitem{bleu}
Kishore Papineni, Salim Roukos, Todd Ward, and Wei-Jing Zhu.
\newblock Bleu: A method for automatic evaluation of machine translation.
\newblock In {\em Annual Meeting of the Association for Computational Linguistics (ACL)}, pages 311--318, 2002.

\bibitem{pearce2022asleep}
Hammond Pearce, Baleegh Ahmad, Benjamin Tan, Brendan Dolan-Gavitt, and Ramesh Karri.
\newblock Asleep at the keyboard? assessing the security of github copilot’s code contributions.
\newblock In {\em IEEE Symposium on Security and Privacy (S\&P)}, pages 754--768, 2022.

\bibitem{magecode}
Hung Pham, Huyen Ha, Van Tong, Dung Hoang, Duc Tran, and Tuyen~Ngoc Le.
\newblock Magecode: Machine-generated code detection method using large language models.
\newblock {\em IEEE Access}, 2024.

\bibitem{gemini-2.0-flash}
Sundar Pichai, Demis Hassabis, and Koray Kavukcuoglu.
\newblock Introducing gemini 2.0: our new ai model for the agentic era, 2024.

\bibitem{ramos2003using}
Juan Ramos et~al.
\newblock Using tf-idf to determine word relevance in document queries.
\newblock In {\em Proceedings of the first instructional conference on machine learning}, volume 242, pages 29--48. Citeseer, 2003.

\bibitem{codebleu}
Shuo Ren, Daya Guo, Shuai Lu, Long Zhou, Shujie Liu, Duyu Tang, Neel Sundaresan, Ming Zhou, Ambrosio Blanco, and Shuai Ma.
\newblock Codebleu: a method for automatic evaluation of code synthesis.
\newblock {\em arXiv preprint arXiv:2009.10297}, 2020.

\bibitem{rosenblatt1958perceptron}
Frank Rosenblatt.
\newblock The perceptron: a probabilistic model for information storage and organization in the brain.
\newblock {\em Psychological review}, 65(6):386, 1958.

\bibitem{rossler2018faceforensics}
Andreas R{\"o}ssler, Davide Cozzolino, Luisa Verdoliva, Christian Riess, Justus Thies, and Matthias Nie{\ss}ner.
\newblock Faceforensics: A large-scale video dataset for forgery detection in human faces.
\newblock {\em arXiv preprint arXiv:1803.09179}, 2018.

\bibitem{sadasivan2023paraphrase}
Vinu~Sankar Sadasivan, Aounon Kumar, Sriram Balasubramanian, Wenxiao Wang, and Soheil Feizi.
\newblock Can ai-generated text be reliably detected?
\newblock {\em arXiv preprint arXiv:2303.11156}, 2023.

\bibitem{sauglam2024automated}
Timur Sa{\u{g}}lam, Sebastian Hahner, Larissa Schmid, and Erik Burger.
\newblock Automated detection of ai-obfuscated plagiarism in modeling assignments.
\newblock In {\em International Conference on Software Engineering: Software Engineering Education and Training (ICSE-SEET)}, pages 297--308, 2024.

\bibitem{detectcodegpt}
Yuling Shi, Hongyu Zhang, Chengcheng Wan, and Xiaodong Gu.
\newblock Between lines of code: Unraveling the distinct patterns of machine and human programmers.
\newblock In {\em International Conference on Software Engineering (ICSE)}, pages 51--62, 2025.

\bibitem{sohl2015deep}
Jascha Sohl-Dickstein, Eric Weiss, Niru Maheswaranathan, and Surya Ganguli.
\newblock Deep unsupervised learning using nonequilibrium thermodynamics.
\newblock In {\em International Conference on Machine Learning (ICML)}, pages 2256--2265. pmlr, 2015.

\bibitem{stackoverflow}
{Stack Overflow}.
\newblock {2024 Stack Overflow Developer Survey}, 2024.

\bibitem{stalnaker2024developer}
Trevor Stalnaker, Nathan Wintersgill, Oscar Chaparro, Laura~A Heymann, Massimiliano Di~Penta, Daniel~M German, and Denys Poshyvanyk.
\newblock Developer perspectives on licensing and copyright issues arising from generative ai for coding.
\newblock {\em arXiv preprint arXiv:2411.10877}, 2024.

\bibitem{steponenaite2023plagiarism}
Aiste Steponenaite and Basel Barakat.
\newblock Plagiarism in ai empowered world.
\newblock In {\em International Conference on Human-Computer Interaction}, pages 434--442, 2023.

\bibitem{suh2025howfar}
Hyunjae Suh, Mahan Tafreshipour, Jiawei Li, Adithya Bhattiprolu, and Iftekhar Ahmed.
\newblock An empirical study on automatically detecting ai-generated source code: How far are we?
\newblock In {\em International Conference on Software Engineering (ICSE)}, 2025.

\bibitem{tambon2025bugs}
Florian Tambon, Arghavan Moradi-Dakhel, Amin Nikanjam, Foutse Khomh, Michel~C Desmarais, and Giuliano Antoniol.
\newblock Bugs in large language models generated code: An empirical study.
\newblock {\em Empirical Software Engineering}, 30(3):1--48, 2025.

\bibitem{tihanyi2025secure}
Norbert Tihanyi, Tamas Bisztray, Mohamed~Amine Ferrag, Ridhi Jain, and Lucas~C Cordeiro.
\newblock How secure is ai-generated code: a large-scale comparison of large language models.
\newblock {\em Empirical Software Engineering}, 30(2):1--42, 2025.

\bibitem{toth2024llms}
Rebeka T{\'o}th, Tamas Bisztray, and L{\'a}szl{\'o} Erd{\H{o}}di.
\newblock Llms in web development: Evaluating llm-generated php code unveiling vulnerabilities and limitations.
\newblock In {\em International Conference on Computer Safety, Reliability, and Security}, pages 425--437, 2024.

\bibitem{turingbench}
Adaku Uchendu, Zeyu Ma, Thai Le, Rui Zhang, and Dongwon Lee.
\newblock Turingbench: A benchmark environment for turing test in the age of neural text generation.
\newblock In {\em Findings of the Association for Computational Linguistics: EMNLP 2021}, pages 2001--2016, 2021.

\bibitem{vaidya2023critical}
Jaideep Vaidya and Hafiz Asif.
\newblock A critical look at ai-generate software: Coding with the new ai tools is both irresistible and dangerous.
\newblock {\em IEEE Spectrum}, 60(7):34--39, 2023.

\bibitem{vaswani2017attention}
Ashish Vaswani, Noam Shazeer, Niki Parmar, Jakob Uszkoreit, Llion Jones, Aidan~N Gomez, {\L}ukasz Kaiser, and Illia Polosukhin.
\newblock Attention is all you need.
\newblock {\em Advances in Neural Information Processing Systems (NeurIPS)}, 30, 2017.

\bibitem{wang2024your}
Jiexin Wang, Xitong Luo, Liuwen Cao, Hongkui He, Hailin Huang, Jiayuan Xie, Adam Jatowt, and Yi~Cai.
\newblock Is your ai-generated code really safe? evaluating large language models on secure code generation with codeseceval.
\newblock {\em arXiv preprint arXiv:2407.02395}, 2024.

\bibitem{openhands}
Xingyao Wang, Boxuan Li, Yufan Song, Frank~F Xu, Xiangru Tang, Mingchen Zhuge, Jiayi Pan, Yueqi Song, Bowen Li, Jaskirat Singh, et~al.
\newblock Openhands: An open platform for ai software developers as generalist agents.
\newblock In {\em International Conference on Learning Representations (ICLR)}, 2025.

\bibitem{wang2023codet5+}
Yue Wang, Hung Le, Akhilesh Gotmare, Nghi Bui, Junnan Li, and Steven Hoi.
\newblock Codet5+: Open code large language models for code understanding and generation.
\newblock In {\em Conference on Empirical Methods in Natural Language Processing (EMNLP)}, pages 1069--1088, 2023.

\bibitem{xiao2022new}
Yunkai Xiao, Soumyadeep Chatterjee, and Edward Gehringer.
\newblock A new era of plagiarism the danger of cheating using ai.
\newblock In {\em International Conference on Information Technology Based Higher Education and Training (ITHET)}, pages 1--6, 2022.

\bibitem{xu2024llms}
Jialiang Xu, Shenglan Li, Zhaozhuo Xu, and Denghui Zhang.
\newblock Do llms know to respect copyright notice?
\newblock In {\em Conference on Empirical Methods in Natural Language Processing (EMNLP)}, pages 20604--20619, 2024.

\bibitem{llmgcode}
Jinwei Xu, He~Zhang, Yanjin Yang, Zeru Cheng, Jun Lyu, Bohan Liu, Xin Zhou, Lanxin Yang, Alberto Bacchelli, Yin~Kia Chiam, et~al.
\newblock Investigating efficacy of perplexity in detecting llm-generated code.
\newblock {\em arXiv preprint arXiv:2412.16525}, 2024.

\bibitem{xu2024licoeval}
Weiwei Xu, Kai Gao, Hao He, and Minghui Zhou.
\newblock Licoeval: Evaluating llms on license compliance in code generation.
\newblock {\em arXiv preprint arXiv:2408.02487}, 2024.

\bibitem{xu2025distinguishing}
Xiaodan Xu, Chao Ni, Xinrong Guo, Shaoxuan Liu, Xiaoya Wang, Kui Liu, and Xiaohu Yang.
\newblock Distinguishing llm-generated from human-written code by contrastive learning.
\newblock {\em ACM Transactions on Software Engineering and Methodology}, 34(4):1--31, 2025.

\bibitem{xu2024detecting}
Zhenyu Xu and Victor~S Sheng.
\newblock Detecting ai-generated code assignments using perplexity of large language models.
\newblock In {\em AAAI Conference on Artificial Intelligence (AAAI)}, volume~38, pages 23155--23162, 2024.

\bibitem{yan2024llm}
Shenao Yan, Shen Wang, Yue Duan, Hanbin Hong, Kiho Lee, Doowon Kim, and Yuan Hong.
\newblock An $\{$LLM-Assisted$\}$$\{$Easy-to-Trigger$\}$ backdoor attack on code completion models: Injecting disguised vulnerabilities against strong detection.
\newblock In {\em USENIX Security Symposium (USENIX Security)}, pages 1795--1812, 2024.

\bibitem{gpt4code}
Xianjun Yang, Kexun Zhang, Haifeng Chen, Linda Petzold, William~Yang Wang, and Wei Cheng.
\newblock Zero-shot detection of machine-generated codes.
\newblock {\em arXiv preprint arXiv:2310.05103}, 2023.

\bibitem{ye2025uncovering}
Tong Ye, Yangkai Du, Tengfei Ma, Lingfei Wu, Xuhong Zhang, Shouling Ji, and Wenhai Wang.
\newblock Uncovering llm-generated code: A zero-shot synthetic code detector via code rewriting.
\newblock In {\em AAAI Conference on Artificial Intelligence (AAAI)}, volume~39, pages 968--976, 2025.

\bibitem{yu2023codeipprompt}
Zhiyuan Yu, Yuhao Wu, Ning Zhang, Chenguang Wang, Yevgeniy Vorobeychik, and Chaowei Xiao.
\newblock Codeipprompt: intellectual property infringement assessment of code language models.
\newblock In {\em International Conference on Machine Learning (ICML)}, pages 40373--40389, 2023.

\bibitem{zeng2025inducing}
Binqi Zeng, Quan Zhang, Chijin Zhou, Gwihwan Go, Yu~Jiang, and Heyuan Shi.
\newblock Inducing vulnerable code generation in llm coding assistants.
\newblock {\em arXiv preprint arXiv:2504.15867}, 2025.

\bibitem{zhang2023well}
Ying Zhang, Wenjia Song, Zhengjie Ji, Na~Meng, et~al.
\newblock How well does llm generate security tests?
\newblock {\em arXiv preprint arXiv:2310.00710}, 2023.

\bibitem{zhu2023genimage}
Mingjian Zhu, Hanting Chen, Qiangyu Yan, Xudong Huang, Guanyu Lin, Wei Li, Zhijun Tu, Hailin Hu, Jie Hu, and Yunhe Wang.
\newblock Genimage: A million-scale benchmark for detecting ai-generated image.
\newblock {\em Advances in Neural Information Processing Systems (NeurIPS)}, 36:77771--77782, 2023.

\bibitem{zi2020wilddeepfake}
Bojia Zi, Minghao Chang, Jingjing Chen, Xingjun Ma, and Yu-Gang Jiang.
\newblock Wilddeepfake: A challenging real-world dataset for deepfake detection.
\newblock In {\em Proceedings of the 28th ACM international conference on multimedia}, pages 2382--2390, 2020.

\end{thebibliography}
\medskip
\bibliographystyle{plain}
}

\newpage
\appendix
To further support and validate our \sys benchmark, we provide the following supplementary materials:
\begin{itemize}
    \item \autoref{appendix:generative model}: Detailed descriptions of the production-level LLMs used in \sys and their corresponding generation settings.
    \item \autoref{appendix:code summarization}: Prompts used in the code summarization phase and representative examples.
    \item \autoref{appendix:code generation}: Prompts used in the code generation phase and representative examples.
    \item \autoref{appendix:code paraphrase}: Domain-specific transformation rules, prompts, used in the code paraphrasing phase with representative examples.
    \item \autoref{appendix:data statistics}: Comprehensive statistics and distributions of the \sys dataset.
    \item \autoref{appendix:baseline}: Detailed descriptions of the baseline detectors included in our evaluation.
    \item \autoref{appendix:tpr}: Supplementary results based on \textit{TPR@FPR} metrics.
    \item \autoref{appendix:results}: Extended and detailed experimental results across all evaluation settings.
    \item \autoref{appendix:limitation}: Additional discussion on the limitations and future improvement directions.
\end{itemize}

\section{Details of Generative Models and Generation Settings}
\label{appendix:generative model}
\begin{table}[h]
    \centering
    \caption{Detailed configurations of the production-level LLMs used in \sys.}
    \resizebox{\linewidth}{!}{
    \begin{tabular}{llll}
    \toprule
    \textbf{LLM Name} & \textbf{API / Model Path} & \textbf{Hyper-Parameter} \\
    \otoprule
    Claude-3.5-Haiku~\cite{claude-3.5-haiku} & \texttt{Anthropic/claude-3-5-haiku-20241022} & temperature = 1.0 \\
    GPT-4o-mini~\cite{gpt-4o-mini} & \texttt{OpenAI/gpt-4o-mini-2024-07-18} & temperature = 1.0 \\
    GPT-o3-mini~\cite{o3-mini} & \texttt{OpenAI/o3-mini-2025-01-31} & \makecell[l]{temperature = 1.0 \\ reasoning\_effort = medium} \\
    Gemini-2.0-Flash~\cite{gemini-2.0-flash} & \texttt{Google/gemini-2.0-flash} & temperature = 1.0 \\
    Gemini-2.0-Flash-Thinking~\cite{gemini-2.0-flash-thinking} & \texttt{Google/gemini-2.0-flash-thinking-exp-01-21} & temperature = 1.0 \\
    Gemini-2.0-Pro~\cite{gemini-2.0-pro} & \texttt{Google/gemini-2.0-pro-exp-02-05} & temperature = 1.0 \\
    DeepSeek-V3~\cite{deepseek-v3} & \texttt{deepseek-ai/DeepSeek-V3} & temperature = 1.0 \\
    DeepSeek-R1~\cite{deepseek-r1} & \texttt{deepseek-ai/DeepSeek-R1} & temperature = 1.0 \\
    Llama-3.3-70B~\cite{llama-3.3-70b} & \texttt{meta-llama/Llama-3.3-70B-Instruct} & temperature = 0.6 \\
    Qwen-2.5-Coder-32B~\cite{qwen-2.5-coder-32b} & \texttt{Qwen/Qwen2.5-Coder-32B-Instruct} & temperature = 0.7 \\
    \bottomrule
    \end{tabular}
    }
    \label{tab:llm config}
\end{table}
In \sys, we adopt ten widely used production-level LLMs from six leading AI companies, including three reasoning models. Detailed configurations and generation settings for these models are presented in \autoref{tab:llm config}. For key generation hyper-parameters such as \texttt{temperature} and \texttt{reasoning\_effort}, we use either default values or officially recommended settings to reflect realistic usage. Importantly, we avoid setting \texttt{temperature} to zero, as doing so would produce overly deterministic outputs that are easier to detect. Instead, we adopt general-purpose settings for high-quality while more diverse and less predictable code generation.

\section{Additional Details of AI Code Summarization}
\label{appendix:code summarization}
To generate high-quality and representative summaries that comprehensively describe the characteristics of a code sample while preventing the leakage of concrete code, we design a structured summarization prompt covering eight key aspects. We then prompt the LLMs to act as summarizers, generating summaries based on the input code file using this carefully crafted prompt. The full summarization prompt used in \sys is as follows:

\begin{tcolorbox}[fancyprompt, title={\faLightbulb\ Summarization Prompt}]
\footnotesize
Analyze the provided code snippet and generate a concise and informative description of its functionality, purpose, and design. Avoid directly including or mirroring the given code. Focus on abstracting the logic, functionality, and intent.
\newline\newline
Follow the output format:
\begin{enumerate}[leftmargin=*]
    \item\textbf{Purpose:} A high-level summary of what the code is intended to achieve.
    \item\textbf{Functionality:} Describe the main tasks performed by the code, including inputs, outputs, and their roles, without referencing exact code or variable names.
    \item\textbf{Logic Overview:} Explain the key logic, algorithms, or patterns conceptually, avoiding specific code structures or syntax.
    \item\textbf{Key Features:} Highlight unique approaches or techniques without mentioning explicit implementation details.
    \item\textbf{Key Libraries:} Include the key libraries used in the code snippet.
    \item\textbf{Key Functions:} Include the key function names used in the code snippet.
    \item\textbf{Key Class, Structure, Variable:} Include the key class, structure, and variable names used in the code snippet.
    \item\textbf{Contextual Notes (optional):} Any assumptions, dependencies, or inferred background information.
\end{enumerate}

\end{tcolorbox}

Below, we present a concrete Python code example with the summary generated by GPT-o3-mini~\cite{o3-mini}, illustrating the consistency and descriptive quality of the summarization process in \sys.

\begin{tcolorbox}[fancyprompt, title=\faFileCode\ Python Summary Example]
\footnotesize
\textbf{Human-Written Code}

\begin{lstlisting}[language=Python,basicstyle=\scriptsize\ttfamily,numbers=none]
import logging
import voluptuous as vol
from homeassistant.components.binary_sensor import
    (BinarySensorDevice, PLATFORM_SCHEMA, SENSOR_CLASSES_SCHEMA)
from homeassistant.components import enocean
from homeassistant.const import (CONF_NAME, CONF_ID, CONF_SENSOR_CLASS)
import homeassistant.helpers.config_validation as cv

_LOGGER = logging.getLogger(__name__)
DEPENDENCIES = ['enocean']
DEFAULT_NAME = 'EnOcean binary sensor'

PLATFORM_SCHEMA = PLATFORM_SCHEMA.extend({
    ...
})

def setup_platform(hass, config, add_devices, discovery_info=None):
    ...
class EnOceanBinarySensor(enocean.EnOceanDevice, BinarySensorDevice):
    def __init__(self, dev_id, devname, sensor_class):
        ...
    @property
    def name(self):
        ...
    @property
    def sensor_class(self):
        ...
    def value_changed(self, value, value2):
        ...
...

\end{lstlisting}

\tcblower

\footnotesize
\textbf{GPT-o3-mini Generated Summary}
\scriptsize
\begin{enumerate}[leftmargin=*]
    \item\textbf{Purpose:} The code integrates binary sensors that communicate using an EnOcean protocol into a larger home automation framework. It enables sensor events to be received, processed, and dispatched within the automation system.
    \item\textbf{Functionality:} The implementation defines a platform component for binary sensors. It validates configuration inputs, creates ...
    \item\textbf{Logic Overview:} The code establishes a configuration schema to ensure correct setup parameters. It then creates a ...
    \item\textbf{Key Features:} Validates configuration using a dedicated schema to ensure data consistency. Leverages inheritance to ...
    \item\textbf{Key Libraries:} Logging, Voluptuous for configuration validation, ...
    \item\textbf{Key Functions:} ... An event handler method that processes changes in sensor values and triggers events ...
    \item\textbf{Key Class, Structure, Variable:} A sensor class that ... , Variables for device identification, sensor naming, ...
    \item\textbf{Contextual Notes:} The design assumes integration within a home automation ecosystem where components communicate via ...
\end{enumerate}

\end{tcolorbox}

For more examples and summaries across different programming languages and generative models, please refer to our \sys dataset.

\section{Additional Details of AI Code Generation}
\label{appendix:code generation}
After generating a comprehensive yet concise summary of the human-written code, we prompt LLMs to produce corresponding AI-generated versions. To preserve structural consistency, we explicitly include the statistical properties (\eg, line of code and character length) of the original code in the prompt. The complete code generation prompt used in \sys is presented below:
\begin{tcolorbox}[fancyprompt, title={\faLightbulb\ Code Generation Prompt}]
\footnotesize
Generate a \texttt{\{Language\}} code snippet with around \texttt{\{Line\_of\_Code\}} lines and \texttt{\{Char\_Length\}} character length following the description. Only output the code, no any other information.
\newline\newline
\texttt{\{Summary\}}
\end{tcolorbox}
The generated code is then passed through a rule-based inspector, as described in \autoref{subsection:benchmark construction}, to filter out overly short, overly long, or null outputs. Below, we show a Python code example generated by GPT-o3-mini~\cite{o3-mini}, based on the summary presented in \autoref{appendix:code summarization}:
\begin{tcolorbox}[fancyprompt, title=\faFileCode\ AI-Generated Python Example]
\footnotesize
\textbf{GPT-o3-mini Code Generation}

\begin{lstlisting}[language=Python,basicstyle=\scriptsize\ttfamily,numbers=none]
import logging
import voluptuous as vol
from homeassistant.const import CONF_NAME, CONF_DEVICE_ID
...
DOMAIN = "enocean_binary_sensor"
CONF_SENSOR_TYPE = "sensor_type"
...
class EventBus:
    def fire(self, event_type, event_data):
        ...
class EnOceanDevice:
    def __init__(self, device_id):
        ...
    def send_message(self, message):
        ...
...
class BinarySensorDevice:
    ...
class EnOceanBinarySensor(EnOceanDevice, BinarySensorDevice):
    ...
...
if __name__ == "__main__":
    logging.basicConfig(level=logging.DEBUG)
    hass = Hass()
    config = {
        CONF_DEVICE_ID: "enocean_001",
        ...
    }
    setup_platform(hass, config)
\end{lstlisting}

\end{tcolorbox}
We observe that though the detailed implementation of the AI-generated code differs from the human-written version, it employs the same key libraries and similar variables to achieve comparable functionality, illustrating the quality of the AI-generated code. More examples are presented in \sys dataset.

\section{Additional Details of AI Code Paraphrasing}
\label{appendix:code paraphrase}
We further evaluate the robustness of detectors under adversarial paraphrasing, a prevalent and practical challenge in real-world applications. Our adversarial setup is grounded in prompt-based paraphrasing, which represents one of the most accessible and commonly used evasion strategies in practice. Unlike prior works on natural language paraphrasing~\cite{krishna2024paraphrasing,sadasivan2023paraphrase}, which rely on generic and concise prompts, we propose a code-specific paraphrasing prompt tailored to the programming domain. This prompt incorporates seven transformation strategies specifically designed for source code. The full prompt is provided below:
\begin{tcolorbox}[fancyprompt, title={\faLightbulb\ Paraphrasing Prompt}]
\footnotesize
Transform the given code to bypass AI-generated code detectors by mimicking human coding styles. Make structural, stylistic, and naming changes while preserving exact functionality. Apply the following:
\begin{enumerate}[leftmargin=*]
    \item Rename variables, functions, and classes with meaningful, human-like names.
    \item Adjust formatting (indentation, spacing, line breaks) and reorganize imports logically.
    \item Rewrite logic with equivalent constructs (\eg, refactor loops, conditionals).
    \item Replace common standard library functions with alternatives or custom helpers.
    \item Vary expressions (\eg, `a + b' → `sum([a, b])') and simplify inline logic into reusable functions.
    \item Transform literals (\eg, single vs. double quotes, `10' → `0xA') and reorder independent code blocks.
    \item Add subtle redundancies (\eg, intermediate variables) to resemble human-written code.
\end{enumerate}

Keep the line of code and character length similar to the original code and only output the transformed code. No any other additional text and information.
\end{tcolorbox}
We then apply the paraphrasing prompt to the AI-generated code sample and obtain its paraphrased version. Below, we present the paraphrased code example produced by GPT-o3-mini~\cite{o3-mini}, based on the AI-generated example provided in \autoref{appendix:code generation}.
\begin{tcolorbox}[fancyprompt, title=\faFileCode\ AI-Paraphrased Python Example]
\footnotesize
\textbf{GPT-o3-mini Paraphrased Code}

\begin{lstlisting}[language=Python,basicstyle=\scriptsize\ttfamily,numbers=none]
import logging
import voluptuous as schema_validator

from homeassistant.const import CONF_NAME, CONF_DEVICE_ID
import homeassistant.helpers.config_validation as config_val

logger = logging.getLogger(__name__)
SENSOR_DOMAIN = 'enocean_binary_sensor'
KEY_SENSOR_KIND = 'sensor_type'
...
SensorConfigSchema = schema_validator.Schema({
    ...
})

class SimulatedEventBus:
    def trigger(self, event_category, particulars):
        ...
    ...

class BaseEnOceanDevice:
    def __init__(self, dev_identifier):
        ...
    def dispatch_message(self, msg_payload):
        ...
class BasicBinarySensor:
    ...
class EnOceanSensorClient(BaseEnOceanDevice, BasicBinarySensor):
    ...
...
if __name__ == '__main__':
    logging.basicConfig(level=logging.DEBUG)
    simulated_hass = SimulatedHomeAssistant()
    input_configuration = {
        ...
    }
    initialize_platform(simulated_hass, input_configuration)

\end{lstlisting}

\end{tcolorbox}
Compared to the original AI-generated code, the paraphrased version uses different aliases for imported libraries, introduces redundant classes and variables, and modifies function and class names with different implementations, while preserving the overall program functionality. Additional examples can be found in the full \sys{} dataset.

\section{Additional Statistics of \sys's Dataset}
\label{appendix:data statistics}
\begin{table}[t]
\centering
\caption{\sys's data quantity statistics across different LLMs and programming languages.}
\resizebox{\linewidth}{!}{
\begin{tabular}{lccccccccccc}
\toprule
\textbf{LLM}                               & \textbf{Paraphrase}    & \textbf{Python} & \textbf{Java} & \textbf{JavaScript} & \textbf{C++}  & \textbf{C}    & \textbf{C\#} & \textbf{Go}   & \textbf{Ruby} & \textbf{PHP}  & \textbf{HTML} \\
\otoprule
Human & \xmark &    1,000   & 1,000 & 1,000       & 1,000 & 1,000 & 1,000   & 1,000 & 1,000 & 1,000 & 1,000 \\
\midrule
\multirow{2}{*}{Claude-3.5-Haiku} & \xmark & 1,000   & 1,000 & 1,000       & 1,000 & 1,000 & 1,000   & 1,000 & 1,000 & 1,000 & 1,000 \\
& \cmark             & 1,000   & 1,000 & 1,000       & 1,000 & 1,000 & 1,000   & 1,000 & 1,000 & 1,000 & 1,000 \\
\midrule
\multirow{2}{*}{DeepSeek-R1} & \xmark & 1,000   & 1,000 & 1,000       & 1,000 & 1,000 & 1,000   & 1,000 & 1,000 & 1,000 & 1,000 \\
& \cmark             & 1,000   & 1,000 & 1,000       & 999  & 1,000 & 999    & 1,000 & 1,000 & 1,000 & 1,000 \\
\midrule
\multirow{2}{*}{DeepSeek-V3} & \xmark & 1,000   & 1,000 & 1,000       & 1,000 & 1,000 & 1,000   & 1,000 & 1,000 & 1,000 & 1,000 \\
& \cmark             & 1,000   & 1,000 & 1,000       & 1,000 & 1,000 & 1,000   & 1,000 & 1,000 & 1,000 & 1,000 \\
\midrule
\multirow{2}{*}{Gemini-2.0-Flash} & \xmark & 1,000   & 1,000 & 1,000       & 1,000 & 1,000 & 1,000   & 1,000 & 1,000 & 1,000 & 1,000 \\
& \cmark             & 1,000   & 1,000 & 1,000       & 1,000 & 1,000 & 1,000   & 1,000 & 1,000 & 1,000 & 1,000 \\
\midrule
\multirow{2}{*}{Gemini-2.0-Flash-Thinking} & \xmark & 1,000   & 1,000 & 1,000       & 1,000 & 1,000 & 1,000   & 1,000 & 1,000 & 1,000 & 1,000 \\
& \cmark             & 1,000   & 1,000 & 1,000       & 1,000 & 1,000 & 1,000   & 1,000 & 1,000 & 1,000 & 1,000 \\
\midrule
\multirow{2}{*}{Gemini-2.0-Pro} & \xmark & 1,000   & 1,000 & 1,000       & 1,000 & 1,000 & 998    & 1,000 & 1,000 & 998  & 999  \\
& \cmark             & 1,000   & 1,000 & 1,000       & 1,000 & 1,000 & 998    & 1,000 & 1,000 & 998  & 999  \\
\midrule
\multirow{2}{*}{GPT-4o-mini} & \xmark & 1,000   & 1,000 & 1,000       & 1,000 & 1,000 & 1,000   & 1,000 & 1,000 & 1,000 & 1,000 \\
& \cmark             & 1,000   & 1,000 & 1,000       & 1,000 & 1,000 & 1,000   & 1,000 & 1,000 & 1,000 & 1,000 \\
\midrule
\multirow{2}{*}{Llama-3.3-70B} & \xmark & 1,000   & 1,000 & 1,000       & 1,000 & 1,000 & 1,000   & 1,000 & 1,000 & 1,000 & 1,000 \\
& \cmark             & 1,000   & 1,000 & 1,000       & 1,000 & 1,000 & 1,000   & 1,000 & 1,000 & 1,000 & 1,000 \\
\midrule
\multirow{2}{*}{GPT-o3-mini} & \xmark &   1,000   & 1,000 & 1,000       & 1,000 & 1,000 & 1,000   & 1,000 & 1,000 & 1,000 & 1,000 \\
& \cmark             & 1,000   & 1,000 & 1,000       & 1,000 & 1,000 & 1,000   & 1,000 & 1,000 & 1,000 & 1,000 \\
\midrule
\multirow{2}{*}{Qwen-2.5-Coder-32B} & \xmark & 1,000   & 1,000 & 1,000       & 1,000 & 1,000 & 1,000   & 1,000 & 1,000 & 1,000 & 1,000 \\
& \cmark             & 1,000   & 1,000 & 1,000       & 1,000 & 1,000 & 1,000   & 1,000 & 1,000 & 1,000 & 1,000\\
\bottomrule
\end{tabular}
}
\label{tab:data statistic}
\end{table}
In \autoref{sec:benchmark statistics}, we present the detailed data quality statistics of the \sys dataset across eight metrics. In this section, we further provide data quantity statistics, as shown in \autoref{tab:data statistic}. For both human-written code and most AI-generated code, we collect or craft 1{,}000 samples per programming language (700 for training and 300 for test). However, some LLMs do not achieve this target for specific languages --- e.g., Gemini-2.0-Pro~\cite{gemini-2.0-pro} on C\# --- due to generation refusals caused by the model's output filtering policies. Despite these occasional omissions, the overall quality of \sys's dataset remain unaffected.

\section{Additional Details of Baseline Detectors}
\label{appendix:baseline}
In this section, we provide additional introduction and implementation details for each of the ten baseline detectors evaluated in \sys.

\parab{\textit{LogRank}~\cite{gltr} \& \textit{Entropy}~\cite{entropy}.}
These two baseline detectors represent classic zero-shot detection approaches that rely on pretrained LLMs. The underlying intuition is that LLMs are more familiar with AI-generated text or code, resulting in lower token-level log-rank or entropy values compared to human-written content. Both methods compute the average token-level statistic (log-rank or entropy) over the input, which is then used as the detection score. In \sys, we implement these detectors using the state-of-the-art open-source pretrained model \textit{Llama-3.2-3B-Instruct}\footnote{\url{https://huggingface.co/meta-llama/Llama-3.2-3B-Instruct}} as the scoring backbone.

\parab{\textit{Binoculars}~\cite{binoculars}.}
\textit{Binoculars} is a state-of-the-art zero-shot detector based on the insight that AI-generated text or code tends to receive more consistent scores across different LLMs than human-written content. To exploit this property, the method feeds the input simultaneously into two distinct LLMs and computes a novel \textit{cross-perplexity} metric as the detection score. In \sys, we adopt the official implementation\footnote{\url{https://github.com/ahans30/Binoculars}} of \textit{Binoculars} to ensure reproducibility and optimized performance.

\parab{\textit{Embed-Code}~\cite{suh2025howfar} \& \textit{Embed-AST}~\cite{suh2025howfar}.} These two embedding-based methods leverage pretrained code embedding models to extract semantic representations of entire code files. \textit{Embed-Code} encodes the raw source code directly, while \textit{Embed-AST} first parses the code into its abstract syntax tree (AST) using \textit{tree-sitter}\footnote{\url{https://github.com/tree-sitter/tree-sitter}}, and then encodes the AST. The embeddings are then passed to a supervised classifier for detection. In \sys, we employ the latest \textit{CodeXEmbed-2B}~\cite{codexembed} model as the embedding model and use a \textit{Random Forest}~\cite{randomforest} classifier as the downstream detector.

\parab{\textit{GPTSniffer}~\cite{gptsniffer-arxiv,gptsniffer-journal}.}
\textit{GPTSniffer} is a state-of-the-art fine-tuning-based detector that leverages the code-related capability of \textit{CodeBERT}~\cite{codebert}. It is fine-tuned on a labeled dataset consisting of both human-written and AI-generated code samples, and evaluated on unseen test data. In \sys, we adopt training hyperparameters consistent with prior work~\cite{codetm4}: 5 training epochs, a learning rate of 3e-4, weight decay of 1e-3, and a warmup ratio of 0.1. We train \textit{GPTSniffer} on \sys's training set and evaluate on \sys's test set.

\parab{\textit{CodeT5+}~\cite{wang2023codet5+} \& \textit{RoBERTa}~\cite{roberta}.}
These two fine-tuning-based detectors follow the same training pipeline as \textit{GPTSniffer}, but utilize different backbone models: the latest \textit{CodeT5+}~\cite{wang2023codet5+} and the classic \textit{RoBERTa}~\cite{roberta}. In \sys, we use the same training hyperparameters and evaluation settings as those employed for \textit{GPTSniffer} to ensure a fair comparison.

\parab{\textit{Raidar}~\cite{raidar}.}
\textit{Raidar} is based on the observation that LLMs tend to modify a greater proportion of human-written content compared to AI-generated content. It hence uses multiple prompts to instruct an LLM to rewrite the input and then computes a set of numerical features (\eg, Bag-of-Words edit distance and Levenshtein score). These features are used to train a downstream classifier as the final detector. In \sys, we adopt the latest \textit{GPT-4.1-nano}\footnote{\url{https://platform.openai.com/docs/models/gpt-4.1-nano}} as the rewriting model, which is stronger than the original \textit{GPT-3.5-Turbo} used in \textit{Raidar}. We also follow the official implementation\footnote{\url{https://github.com/cvlab-columbia/raidarllmdetect}} to extract features and train the detection model.

\parab{\textit{BiScope}~\cite{biscope}.}
\textit{BiScope} is a state-of-the-art detector that leverages a pre-trained LLM to extract bi-directional entropy features, which are then used to train a lightweight downstream classifier. The bi-directional entropy is designed to capture both next-token prediction (forward entropy) and previous-token memorization (backward entropy) from the model’s output logits. In \sys, we use \textit{Llama-3.2-3B-Instruct}\footnote{\url{https://huggingface.co/meta-llama/Llama-3.2-3B-Instruct}} as the feature extractor for \textit{BiScope}, consistent with the scoring model used in \textit{LogRank} and \textit{Entropy}. A \textit{Random Forest}~\cite{randomforest} classifier is employed as the downstream detector.

\section{Evaluation Results of Additional Metrics}
\label{appendix:tpr}

\begin{figure}[t]
    \centering
    \includegraphics[width=1\linewidth]{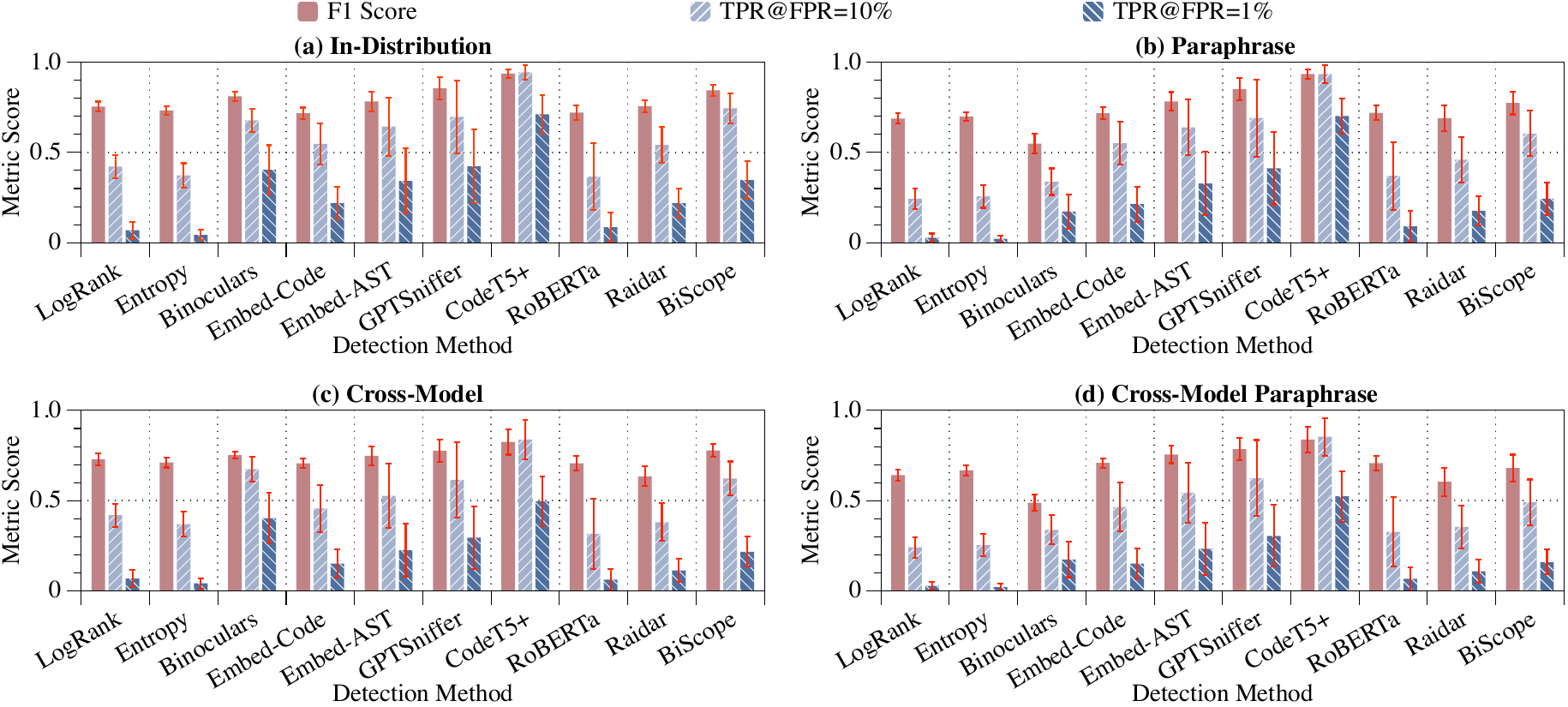}
    \caption{\textbf{Comparison Between Different Evaluation Metrics.} The bar charts illustrate the average F1 scores of baseline detectors on different LLMs across programming languages.}
    \label{fig:metric-comparison}
\end{figure}

The results appear in \autoref{fig:metric-comparison}, where the x-axis lists the detection methods and the y-axis shows their metric values. As before, each bar reflects the mean performance across ten programming languages and ten LLMs, with error bars indicating one standard deviation. The figure is divided into four panels, each corresponding to a different evaluation configuration.
Despite decent F1 scores across the board, all detectors suffer a dramatic drop in true‐positive rate once the false-positive rate is constrained (e.g., TPR@FPR=1\% is generally lower than 0.3), showing that they fail to catch enough positives under realistic, low-alarm requirements and are therefore impractical.

\section{Additional Evaluation Results}
\label{appendix:results}
\begin{figure}
    \centering
    \includegraphics[width=1\linewidth]{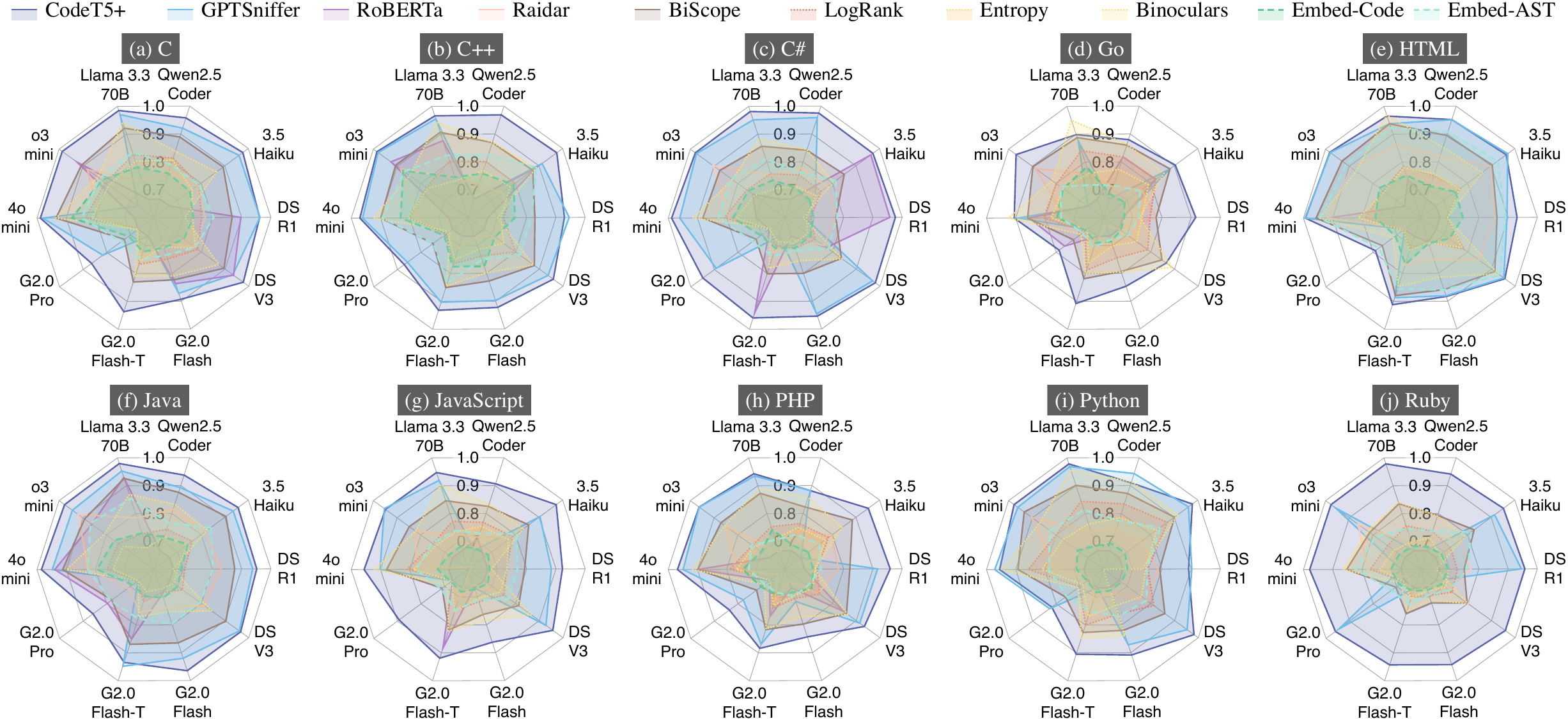}
    \caption{Complete F1 scores of all baseline detectors across various LLMs and programming languages under the in-distribution configuration.}
    \label{fig:id f1 radar}
\end{figure}
\begin{figure}
    \centering
    \includegraphics[width=1\linewidth]{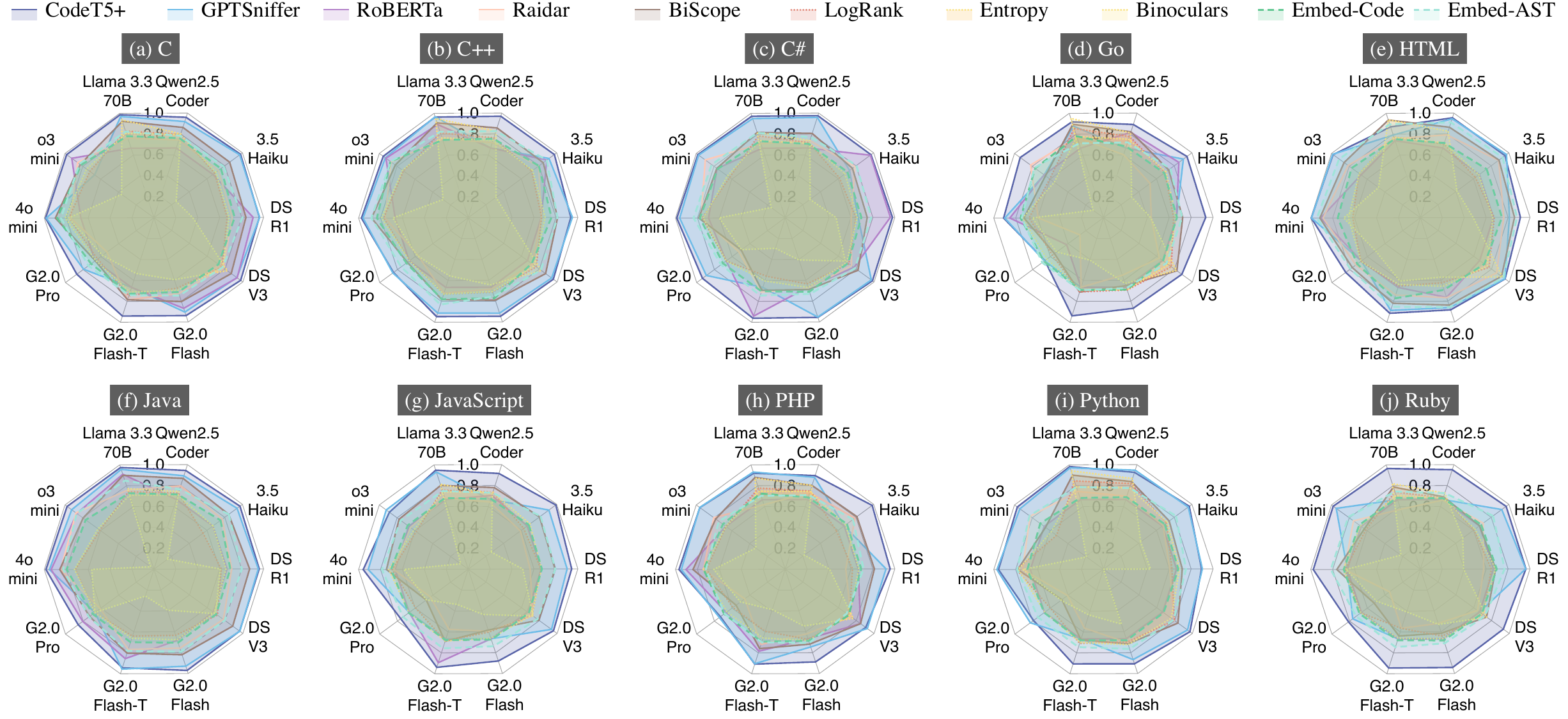}
    \caption{Complete F1 scores of all baseline detectors across various LLMs and programming languages under the paraphrase configuration.}
    \label{fig:paraphrase f1 radar}
\end{figure}
\begin{figure}
    \centering
    \includegraphics[width=1\linewidth]{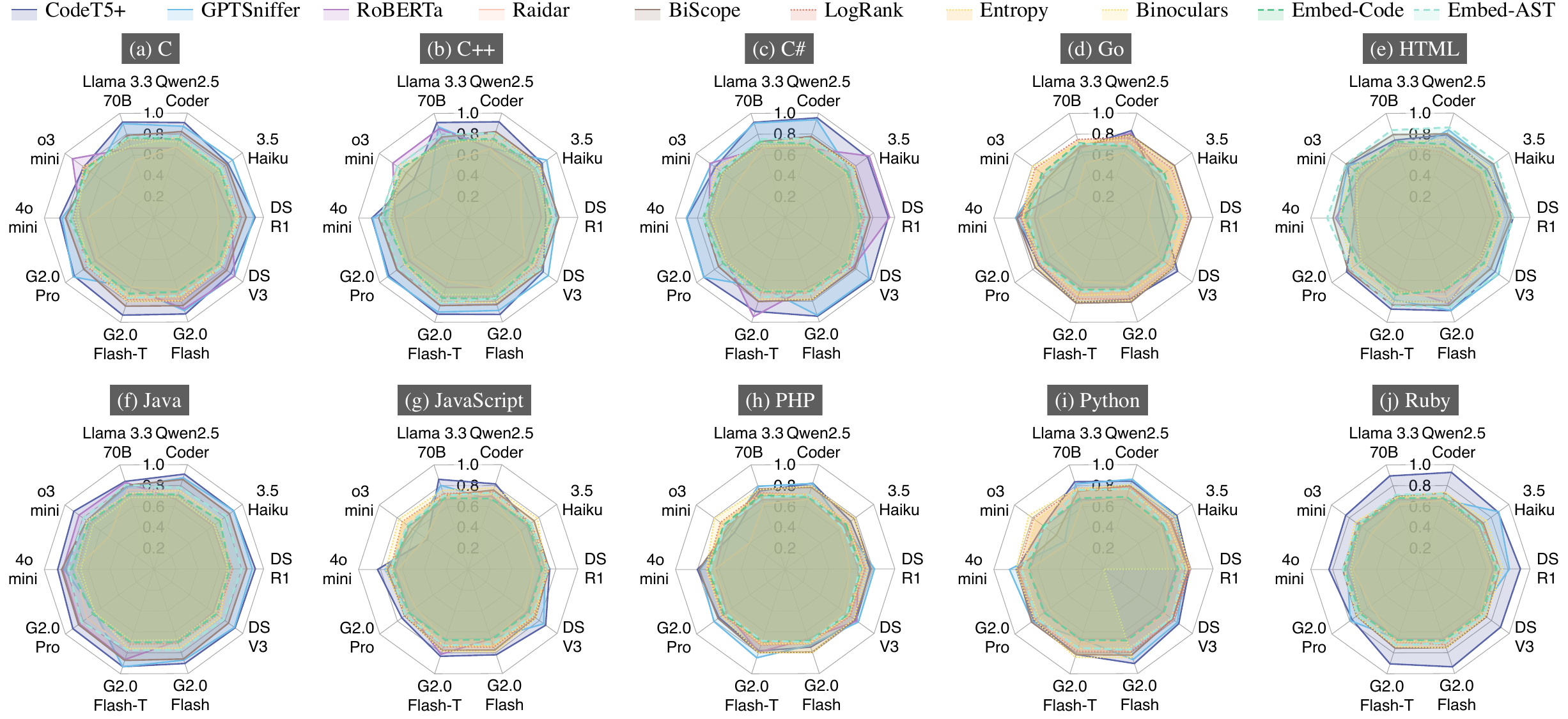}
    \caption{Complete F1 scores of all baseline detectors across various LLMs and programming languages under the cross-model configuration.}
    \label{fig:cm f1 radar}
\end{figure}
\begin{figure}
    \centering
    \includegraphics[width=1\linewidth]{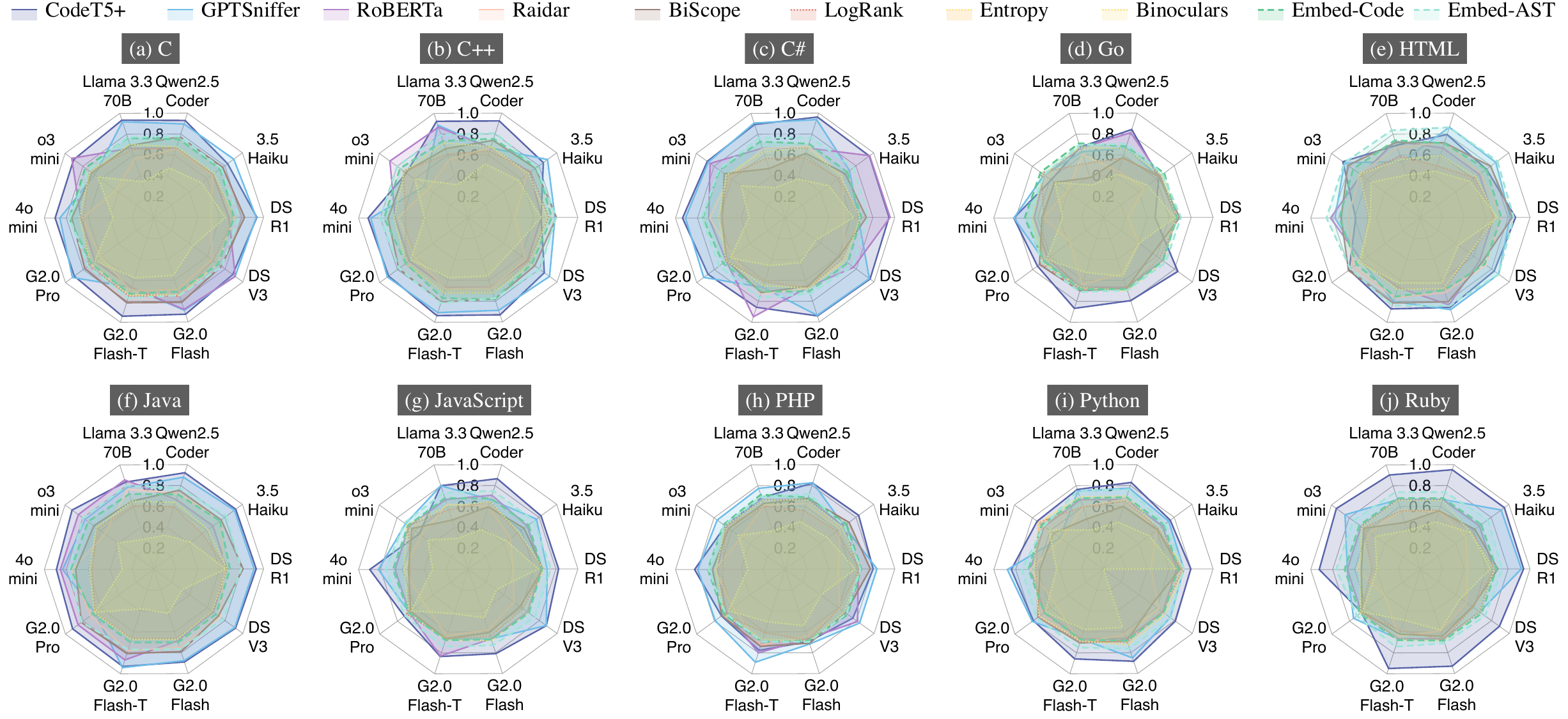}
    \caption{Complete F1 scores of all baseline detectors across various LLMs and programming languages under the cross-model paraphrase configuration.}
    \label{fig:cm paraphrase f1 radar}
\end{figure}

In this section, we present the complete F1 scores of all baseline detectors evaluated across different LLMs and programming languages. Specifically, \autoref{fig:id f1 radar} shows the results under the in-distribution configuration, while \autoref{fig:paraphrase f1 radar} reports the scores under the paraphrase configuration. \autoref{fig:cm f1 radar} illustrates the results under the cross-model configuration, and \autoref{fig:cm paraphrase f1 radar} presents the scores under the cross-model paraphrase configuration.

These comprehensive results are consistent with the trends discussed in \autoref{section:evaluation}, further validating the key findings derived from the \sys evaluation.

\section{Limitations and Future Work}
\label{appendix:limitation}
While \sys represents a significant step toward a more comprehensive evaluation of AI-generated code detectors, several limitations remain and could be addressed in future work.

First, though \sys includes a broad set of programming languages, LLMs, and detectors, it does not exhaustively cover all possibilities. Additional languages, particularly those less commonly used in mainstream software development but still important in specific domains, remain unexplored. Similarly, many emerging LLMs and detection techniques are not included in the current benchmark. Future work could expand \sys to incorporate these newly emerged models and underrepresented languages, enabling broader and more inclusive evaluations.

Second, \sys focuses exclusively on prompt-based paraphrasing attacks in its adversarial setting, given their practicality and prevalence in real-world coding. However, a wider spectrum of adversarial techniques, especially those designed for the natural language domain, could be explored. Future efforts could adapt adversarial strategies against natural language detection to the code domain or propose novel code-specific attack paradigms to more rigorously evaluate detector robustness.

Third, \sys centers on document-level detection where AI-generated code files are fully generated by LLMs. In practice, however, AI coding assistants often generate partial code completions embedded in human-written code. Evaluating detection methods under mixed-authored code with different granularities could be an important direction for future benchmarks and detection methods.

Despite these limitations, \sys advances the field by offering a more comprehensive and realistic evaluation benchmark compared to prior work~\cite{suh2025howfar,pan2024assessing,aigcodeset,magecode,codetm4,llmgcode}. We believe the insights obtained and evaluation platform established by \sys will serve as a strong foundation for developing more robust and generalizable AI-generated code detectors.


\end{document}